\def\bq{\begin{equation}}
\def\eq{\end{equation}}
\def\bqa{\begin{eqnarray}}
\def\eqa{\end{eqnarray}}
\def\bqb{\begin{eqnarray*}}
\def\eqb{\end{eqnarray*}}
\documentstyle[12pt,epsf]{article}
\def\pr#1#2#3{ Phys. Rev. ${\bf{#1}}$ (#2) #3}
\def\prl#1#2#3{ Phys. Rev. Lett. ${\bf{#1}}$ (#2) #3}
\def\pl#1#2#3{ Phys. Lett. ${\bf{#1}}$ (#2) #3}

\def\np#1#2#3{ Nucl. Phys. ${\bf{#1}}$ (#2) #3}
\def\zp#1#2#3{ Z. Phys. ${\bf{#1}}$ (#2) #3}

\def\ie{{\it i.e.\/}}
\def\eg{{\it e.g.\/}}

\def\etal{{\it et.al.\/}}

\global\nulldelimiterspace = 0pt
 
\def\roughly#1{\mathrel{\raise.3ex
    \hbox{$#1$\kern-.75em\lower1ex\hbox{$\sim$}}}}
\def\lsim{\roughly<}
\def\gsim{\roughly>}

\def\L{ {\cal L }}
\def\O{ {\cal O }}
\def\M{ {\cal M }}
\def\R{ {\cal R }}
\def\sw{s_W}
\def\cw{c_W}
\def\swd{s^2_W}
\def\cwd{c^2_W}

\def\mwd{M_W^2}
\def\mw{M_W}
\def\mt{m_t}
\def\mtd{m_t^2}
\def\lw{\lambda_W}

\def\lam{\Lambda_{NP}} 
\def\pvec{\overrightarrow p}
\def\kvec{\overrightarrow k}

\begin{document}
\pagenumbering{arabic}
\thispagestyle{empty}
\def\thefootnote{\fnsymbol{footnote}}
\setcounter{footnote}{1}
 
\begin{flushright}
hep-ph/9606435 \\ 
PM/96-22 \\ THES-TP 96/06 \\
June 1996 \\
 \end{flushright}
\vspace{2cm}
\begin{center}
{\Large\bf Effective lagrangian description of top production and decay
}\footnote{Partially supported by the EC contract CHRX-CT94-0579  and
by Ministry of Education, 
Science and Culture, Japan, under Grant-in-Aid for International 
Scientific Research Program (No.07044097).}
 \vspace{1.5cm}  \\
{\large G.J. Gounaris$^a$, M. Kuroda$^b$, F.M. Renard$^c$}
\vspace {0.5cm}  \\

$^a$Department of Theoretical Physics, University of Thessaloniki,\\
Gr-54006, Thessaloniki, Greece.\\
\vspace{0.2cm}
$^b$ Institute of Physics, Meiji-Gakuin University\\
Yokohama, Japan.\\
\vspace{0.2cm}
$^c$ Physique
Math\'{e}matique et Th\'{e}orique,
UPRES-A 5032\\
Universit\'{e} Montpellier II,
 F-34095 Montpellier Cedex 5.\\

\vspace {1cm}
 
{\bf Abstract}
\end{center}

We propose a rather general description of residual New Physics
(NP)  effects
on the top quark couplings. These effects are described in terms of
20 gauge invariant
$dim=6$ operators involving gauge and Higgs bosons as well as quarks of the
third family. We compute their implications for  the
$\gamma t\bar t$, $Z t \bar t$ and
$tbW$ vertices and study their observability in the process
$e^-e^+ \to t\bar t$ with $t\to bW \to  b\ell^+\nu_\ell$. We 
present results for   the integrated cross section, the
angular distribution and various decay distribution and polarization
asymmetries for NLC energies of $0.5-2~TeV$. Observability limits
are discussed and interpreted in terms of the NP
scales associated to each operator through the unitarity
constraints. The general landscape of the residual
NP effects in the heavy quark and  bosonic
sectors is also  presented.\\
\vspace{0.5cm}
\def\thefootnote{\arabic{footnote}}
\setcounter{footnote}{0}
\clearpage

\section
{Introduction}
\hspace{0.5cm}
It is commonly hoped that the high value of the top mass may  open a
window towards understanding the mass generation mechanism. 
It is then important to look
at the top quark interactions in a very accurate way, searching
for possible departures from universality,
which are somehow associated with the heavy 
$t$-mass; \ie~ 
differences from the properties of the  light
quarks and leptons. This is particularly true, 
if no new particles are lying
within the reach of the contemplate Colliders.
Such top-mass effects
are of course well known
for certain Standard Model (SM) electroweak radiative
corrections. They increase like $m^2_t$; as \eg~
 in the so-called $\delta\rho$ or $\epsilon_1$ parameters
\cite{mt2}. Our search for NP will then concentrate on 
whether there
exist any additional effects, somehow related to the scalar
sector and the large top-mass, $\mt$, which are
beyond those expected in SM. \par

The most intriguing hint towards this kind of NP is provided by
the present situation ~concerning the $Zb\bar b$ vertex. 
This vertex receives a well known SM ~contribution \cite{mt2bb}
proportional to $\mtd$, which does not seem sufficient though, 
to explain the data. Indeed, the experimental
results at LEP1 and
at SLC suggest that the SM top effect in $\epsilon_1$ agrees
with the top mass value found at Fermilab, 
while no corresponding agreement
is observed for $\Gamma(Z\to b\bar b)$ \cite{confbb}. 
Thus, if the present experimental result
on $\Gamma(Z\to b\bar b)$ is correct, it probably indicates the
appearance of a mechanism whose origin must lie beyond SM.
Various types of ideas for this new physics are possible, 
partially ~stemming from the fact that the Left-versus-Right 
structure of
the $Zb\bar b$ vertex is not yet completely
established \cite{bb1, bb, bbother}. \par

An additional
measurement allowing to determine the asymmetry factor $A_b$ is 
required to clarify the $Z\to b \bar b$ situation. 
This can be achieved either by measuring the
forward-backward asymmetry at LEP1, or more directly from the 
measurement at SLC of the polarized forward-backward asymmetry
\cite{confbb}. At present there exists some disagreement between
these two measurements. But in any case, the data seem to
suggest that if an NP effect is present, then it should 
predominantly be affecting the right-handed $Zb\bar b$ amplitude. 
The situation is further 
complicated by the observation of a
(weaker) anomalous effect in the $Zc\bar c$ vertex, which cannot
be associated to an obvious virtual top quark contribution 
and requires a more direct NP source affecting light
quarks also \cite{Zprime}.\par

We assume here, that in the ~foreseeable future, no new
particles will be found, beyond
those present in\footnote{Higgs is an old particle in this
sense.} SM. In such a case, NP could only
appear in the form of residual interactions
generated at a very large scale; \ie~ $\lam \gg M_W$. 
It may turn out
that these residual interactions stem from the
scalar sector and affect only Higgs
and its "partners"; i.e. the fermions coupled most strongly to the scalar 
sector and the gauge bosons.  Under these conditions, 
NP should be described by an effective lagrangian expressed in
terms of $SU(3)_c\times SU(2)\times U(1)$ gauge
invariant $dim=6$ operators, involving  the Higgs together with
$W, Z, \gamma $, the quarks of the third family and the
gluon. To somewhat
restrict the number of such  operators, we impose the constraint
that the quark depending operators should necessarily involve at
least one $t_R$ field \cite{bb}. We do this motivated by the
form of the SM Yukawa couplings in the limit where all fermion
masses, except the top, are neglected.  
In the present work we also impose $CP$ invariance for NP, and
disregard operators which, (after the use of the 
equations of motion), would have rendered 4-quark operators
involving leptons or light quarks of the first two families.\par 

There exist 14 such  CP-conserving "top" operators which have
been classified  in \cite{bb}. In addition, there exist 
 6 CP-conserving purely bosonic
ones, which are "blind" with respect to the LEP1 Physics
and have been studied in \cite {Hag, DeR}.
The indirect constraints on these operators from the LEP1/SLC
measurements (including those from $Z\to b\bar b$),   have been
established in \cite{bb} for the "top" ~operators,
and in \cite{DeR, Hag, bb1} for the bosonic ones. These constraints 
appear to be quite mild, 
calling for a more detail study in a higher energy collider.
Towards this aim, the observable signatures of the bosonic NP operators 
through  the high energy processes
$e^+e^-\to W^+W^-$ \cite{BMT, LEP2TGC}, $e^+e^-\to HZ$ and
$e^+e^-\to H\gamma$ \cite{HZ, HagHiggs} have already been studied.\par

In the present work we are interested in 
the more \underline{direct tests}
of the above operators that will
be provided by the real top production process 
$e^+e^-\to t\bar t$ and the decay  
$t\to Wb \to b l^+ \nu$. 
Such a study should considerably improve the ~existing sensitivity
limits on this kind of NP.
Since we expect the NP effects to be rather small, it is
sufficient for the calculation to restrict ourselves
to the leading contributions. Thus, we either restrict
ourselves to the tree level contribution, whenever this is
non-vanishing or (if it vanishes) retain only the
leading-log 1-loop effect, provided it is enhanced by a power of 
the large top  mass, $\mt$. Under these conditions, box diagrams are
never important in our studies. Thus, we just need to determine
the NP effects on the
$\gamma t\bar t$, $Zt\bar t$ and $tbW$ vertices.  \par

The interesting physical quantities are
the integrated cross section, the density matrix 
of the produced $t$ quark, and various angular distributions. 
We show how the density matrix elements can
be measured through decay distributions with or without $e^{\pm}$-beam
polarization. The sensitivities of the various observables to
each operator are presented and the observability
limits for the associated couplings are established. With the
help of unitarity relations, these
limits are translated into lower bounds on the scale of
corresponding New Physics. Finally we draw the panorama of the 
knowledge that one can
reach on the whole set of top and  bosonic operators through
these tests, as well as through the previous indirect ones.\par

The contents of the paper are the following: In Section 2 we
enumerate  the
$dim=6$ operators used to construct the effective NP lagrangian,
and define their couplings and associated NP scales. In Section 3
we compute the corresponding general $e^+e^-\to t\bar t$ helicity
amplitude and the top density matrix. Section 4 is devoted to 
the $t \to bW \to  b\ell^+\nu_\ell $ decay of the produced top quark. 
In Section 5 we compute the SM 1-loop
$m^2_t$-enhanced contributions to these amplitudes; 
and in Section 6 the
leading NP contributions for all operators considered . The
resulting panorama of residual NP effects  is discussed in Section 7.
Two appendices give details about the computations of the NP
effects and the decay distributions and asymmetries.

\section{The effective lagrangian.}
\hspace{0.5cm}
The list of the $dim=6$, $SU(3)_c \times SU(2)\times
U(1)$ gauge invariant and CP-conserving operators
involving the third family of quarks with at least
one $t_R$ field\footnote{These quark fields are of course
understood as the weak-eigenstate fields. They are related to
the fields creating or absorbing the mass-eigenstates through the
usual unitary transformations leading to the CKM matrix.}, 
together with gauge and
scalar boson fields, has been
established in \cite{bb}. These operators are divided into  two
groups containing four and two quark fields respectively.\\
 
\vspace{0.5cm}
 
1) \underline {Four-quark operators}\\
\bqa
\O_{qt} & = & (\bar q_L t_R)(\bar t_R q_L) \ \ \ , \ \\[0.1cm]
\O^{(8)}_{qt} & = & (\bar q_L \overrightarrow\lambda t_R)
(\bar t_R \overrightarrow\lambda q_L) \ \ \ ,\ \\[0.1cm]
\O_{tt} & = & {1\over2}\, (\bar t_R\gamma_{\mu} t_R)
(\bar t_R\gamma^{\mu} t_R) \ \ \ , \ \\[0.1cm]
\O_{tb} & = & (\bar t_R \gamma_{\mu} t_R)
(\bar b_R\gamma^{\mu} b_R) \ \ \ , \ \\[0.1cm]
\O^{(8)}_{tb} & = & (\bar t_R\gamma_{\mu}\overrightarrow\lambda t_R)
(\bar b_R\gamma^{\mu} \overrightarrow\lambda b_R) \ \ \ , \\[0.1cm]
\O_{qq} & = & (\bar t_R t_L)(\bar b_R b_L) +(\bar t_L t_R)(\bar
b_L b_R)\ \ \nonumber\\
\null & \null & - (\bar t_R b_L)(\bar b_R t_L) - (\bar b_L t_R)(\bar
t_L b_R) \ \ \ , \ \\[0.1cm]
\O^{(8)}_{qq} & = & (\bar t_R \overrightarrow\lambda t_L)
(\bar b_R\overrightarrow\lambda b_L)
+(\bar t_L \overrightarrow\lambda t_R)(\bar b_L
\overrightarrow\lambda  b_R)
\ \nonumber\\
\null & \null &
- (\bar t_R \overrightarrow\lambda b_L)
(\bar b_R \overrightarrow\lambda t_L)
- (\bar b_L \overrightarrow\lambda t_R)(\bar t_L
\overrightarrow\lambda   b_R)
\ \ \  . \
\eqa\\
2) \underline {Two-quark operators.}\\
\bqa
\O_{t1} & = & (\Phi^{\dagger} \Phi)(\bar q_L t_R\widetilde\Phi +\bar t_R
\widetilde \Phi^{\dagger} q_L) \ \ \ ,\ \\[0.1cm]
\O_{t2} & = & i\,\left [ \Phi^{\dagger} (D_{\mu} \Phi)- (D_{\mu}
\Phi^{\dagger})  \Phi \right ]
(\bar t_R\gamma^{\mu} t_R) \ \ \ ,\\[0.1cm]
\O_{t3} & = & i\,( \widetilde \Phi^{\dagger} D_{\mu} \Phi)
(\bar t_R\gamma^{\mu} b_R)-i\, (D_{\mu} \Phi^{\dagger}  \widetilde\Phi)
(\bar b_R\gamma^{\mu} t_R) \ \ \ ,\\[0.1cm]
\O_{D t} &= & (\bar q_L D_{\mu} t_R)D^{\mu} \widetilde \Phi +
D^{\mu}\widetilde \Phi^{\dagger}
((D_{\mu}\bar t_R) ~ q_L) \ \ \ , \\[0.1cm]
\O_{tW\Phi} & = & (\bar q_L \sigma^{\mu\nu}\overrightarrow \tau
t_R) \widetilde \Phi \cdot
\overrightarrow W_{\mu\nu} + \widetilde \Phi^{\dagger}
(\bar t_R \sigma^{\mu\nu}
\overrightarrow \tau q_L) \cdot \overrightarrow W_{\mu\nu}\ \ \
,\\[0.1cm]
\O_{tB\Phi}& = &(\bar q_L \sigma^{\mu\nu} t_R)\widetilde \Phi
B_{\mu\nu} +\widetilde \Phi^{\dagger}(\bar t_R \sigma^{\mu\nu}
 q_L) B_{\mu\nu} \ \ \ ,\\[0.1cm]
\O_{tG\Phi} & = & \left [ (\bar q_L \sigma^{\mu\nu} \lambda^a t_R)
\widetilde \Phi
 + \widetilde \Phi^{\dagger}(\bar t_R \sigma^{\mu\nu}
\lambda^a q_L)\right ] G_{\mu\nu}^a  \ \ \ ,
\eqa
where $\lambda^a ~ (a=1,..., 8)$ are the usual eight colour 
matrices.\par

In the preceding ~formulae the  definitions
\bq 
\Phi=\left( \begin{array}{c}
      i\chi^+ \\
{1\over\sqrt2}(v+H-i\chi^3) \end{array} \right) \ \ \ \ , \ \	
\eq
\bq
D_{\mu}  =  (\partial_\mu + i~ g\prime\,Y B_\mu +
i~ \frac{g}{2} \overrightarrow \tau \cdot \overrightarrow W_\mu  +
i~ \frac{g_s}{2} \overrightarrow \lambda \cdot \overrightarrow G_\mu ) \ \
 , \ 
\eq
are used where $v\simeq 246 ~GeV$, $Y$ is the hypercharge of the
field on which the
covariant derivative acts, and $\overrightarrow \tau$ and  
$\overrightarrow \lambda$ are the isospin and colour
matrices applicable whenever $D_\mu$ acts on isodoublet fermions
and quarks respectively. As already stated, in writing  (1-14),
we have used the equations of motion. If these later equations 
were not used, then  two more operators are met which contain 
additional derivatives \cite{unitop}. \par
 
\vspace{0.5cm}
\noindent 
3) \underline {Bosonic operators.}\\
In addition to the above fermionic $dim=6$ operators, 
NP may also be hiding in purely bosonic
ones. Provided CP invariance is imposed, this kind
of NP is described by 11  $dim=6$ purely bosonic
operators first classified in \cite{Hag}.  
Here, we retain  only  the
six "blind" or "super-blind" ones \cite{bosonic}, which
are not severely constrained by the $Z$-peak  
experiments \cite{HagMS}. They are:
\bqa 
\O_W &= & {1\over3!}\left( \overrightarrow{W}^{\ \ \nu}_\mu\times
  \overrightarrow{W}^{\ \ \lambda}_\nu \right) \cdot
  \overrightarrow{W}^{\ \ \mu}_\lambda \ \ \ ,  \ \ \\[0.1cm]
\O_{W\Phi} & = & i\, (D_\mu \Phi)^\dagger \overrightarrow \tau
\cdot \overrightarrow W^{\mu \nu} (D_\nu \Phi) \ \ \  , \ \
 \\[0.1cm]
\O_{B\Phi} & = & i\, (D_\mu \Phi)^\dagger B^{\mu \nu} (D_\nu
\Phi)\ \ \  . \ \  \\[0.1cm]
\O_{UW} & \equiv & \frac{1}{v^2}\,(\Phi^\dagger \Phi-\frac{v^2}{2})
\, \overrightarrow W^{\mu\nu} \cdot \overrightarrow W_{\mu\nu} \ \
\  ,  \ \ \\[0.1cm]
\O_{UB} & \equiv & \frac{4}{v^2}\,(\Phi^\dagger
 \Phi-\frac{v^2}{2}) B^{\mu\nu} \,B_{\mu\nu} \ \ \  , \ \ \\[0.1cm]
\O_{\Phi 2} & = & 4 ~ \partial_\mu (\Phi^\dagger \Phi)
\partial^\mu (\Phi^\dagger \Phi ) \ \ \  . \ \ \
\eqa
The remaining five operators (called $\O_{DW}, 
\O_{DB}, \O_{BW}, \O_{\Phi 1},
\O_{\Phi 3}$) \cite{Hag} are ignored here, 
since they are either too ~severely
constrained, or they give no ~contribution,
(up to the 1-loop level), to the processes 
we study\footnote{For a possibly less strong constraint on 
$\O_{DW}, \O_{DB}$ see \cite{HagMS}, \cite{clean}}.   \par

The resulting effective lagrangian describing the
residual NP interactions is written as
\bq
\L = \L_t + \L_{bos} \ \ , \
\eq
where the contribution from the 14 ($i=1 ... 14$) "top" operators 
is
\bq
\L_t =  \sum_i { f_i \over m^2_t}\,\O_i \ \ \ , \
\eq
while the ~purely bosonic ones give 
\begin{equation}
{\cal L}_{bos}=\lambda_W{g\over M^2_W}\O_W+f_W{g\over2M^2_W}
\O_{W\Phi}+f_B{g'\over2M^2_W}  \O_{B\Phi}+
d\ {\cal O}_{UW}+{\frac{d_B}4}
\ {\cal O}_{UB}+ \frac{f_{\Phi 2}}{v^2}
{\cal O}_{\Phi 2}\ \ \ \ \ \ .\
\end{equation}
As a whole we have 20 independent operators
that we shall ~occasionally globally label as $\O_{i}$, with 
$i=1, ...  20$. \par
 
To each of the coupling constants $f_i$ (or $\lw$, $d$, $d_B$) appearing in
this lagrangian, a corresponding New Physics scale
 $\Lambda_{NP}$ is associated through the unitarity 
relations established in
\cite{unit, HZ, unitop}. Obviously
$\Lambda_{NP}$ generally depends on the operator $\O_i$ considered.
Thus for the 6 purely bosonic operators
we have found\footnote{The expression for $f_{\Phi2}$ is only valid for
$\lam \gg 3.7TeV$. A more detail discussion is given in \cite{HZ}.}
 \cite{unit, HZ}
\bq
|\lambda_W| \simeq 19~{M^2_W \over \lam^2} \ \ \ \ \ , \ \ \ \ \ \ \
|f_B| \simeq 98~{M^2_W\over\lam^2} \ \ \ \ \ , \ \ \ \ \
 \ |f_W| \simeq 31~{M^2_W \over\lam^2} \ \ \ \ ,\
\eq
\begin{eqnarray}
d & \simeq & \frac{104.5~\left ({\frac{M_W}{\Lambda_{NP}}}
\right )^2} {1+6.5 \left ({\frac{M_W}{\Lambda_{NP}}}\right )} \ \
\mbox{ for } d>0 \ , \nonumber \\
d & \simeq & -~ \frac{104.5~\left ({\frac{M_W}{\Lambda_{NP}}}
\right )^2} {1- 4 \left ({\frac{M_W}{\Lambda_{NP}}}\right )} \ \
\mbox{ for }\  d<0 \ , \  \\
d_B & \simeq & \frac{195.8 ~\left ({\frac{M_W}{\Lambda_{NP}}}
\right )^2} {1+200 \left  ({\frac{M_W}{\Lambda_{NP}}}
\right )^2}\ \
\mbox{ for } d_B>0 \ , \nonumber \\
d_B & \simeq & -~ \frac{195.8 ~\left ({\frac{M_W}{\Lambda_{NP}}}
\right )^2} {1 +50 \left  ({\frac{M_W}{\Lambda_{NP}}}
\right )^2}\ \
\mbox{ for }\  d_B<0 \ , \  \\
|f_{\Phi2}| & \simeq & 75\frac{M^2_W}{\Lambda^2_{NP}} \ \ 
 \ .
\eqa
On the other hand, for the 14 "top" operators, unitarity 
gives\footnote{In the expression for $\O_{tG\Phi}$ we assumed 
$\lam \lsim 10TeV$. Our results are derived by considering
four-body amplitudes at the tree approximation. This may not be
adequate for $\O_{t1}$ which is given by the standard top Yukawa
interaction multiplied by $\Phi^\dagger \Phi$. This problem is
not further investigated though, since $\O_{t1}$ never
contributes to the processes studied here.} \cite{unitop} 
\bqa
|f_{qt}| & \simeq & {16\pi\over3}~
\left ({m^2_t\over \Lambda^2_{NP}}\right ) \ \ ,\\[0.1cm]
|f^{(8)}_{qt}| & \simeq & {9\pi\over\sqrt2}~
\left ({m^2_t\over \Lambda^2_{NP}}\right ) \ \ , \\[0.1cm]
|f_{tt}| & \simeq & 6\pi~ \left ({m^2_t\over \Lambda^2_{NP}}
\right ) \ \ ,\\[0.1cm]
|f_{tb}| & \simeq & 8\pi ~ \left ({m^2_t\over \Lambda^2_{NP}}
\right ) \ \ ,\\[0.1cm]
|f^{(8)}_{tb}| & \simeq & {9\pi\over 2}
\left ({m^2_t\over \Lambda^2_{NP}} \right )\ \ , \\[0.1cm] 
|f_{qq}| & \simeq & {32\pi\over7}~
\left ({m^2_t\over \Lambda^2_{NP}}\right ) \ \ , \\[0.1cm]
 |f^{(8)}_{qq}| & \simeq & 6\pi ~ \left ({m^2_t\over 
\Lambda^2_{NP}} \right ) \ \ , 
\eqa
\bqa
|f_{t1}| & \simeq & {16\pi\over3\sqrt2}~
\left ({m^2_t\over v\Lambda_{NP}} \right ) \ \ , \\[0.1cm]
|f_{t2}| & \simeq & 8\pi\sqrt3 ~
\left ({m^2_t\over \Lambda^2_{NP}}\right ) \ \ , \\[0.1cm]
|f_{t3}| & \simeq & 8\pi\sqrt6 ~
\left ({m^2_t\over \Lambda^2_{NP}} \right )\ \ , \\[0.1cm]
|f_{Dt}| & \simeq & 8.2  ~
\left ({m^2_t\over \Lambda^2_{NP}} \right ) \ \ , \\[0.1cm]
|f_{tW\Phi}| & \simeq & {61.6\over\sqrt{1+645\, {m^2_t\over 
\Lambda^2_{NP}}}}~
\left ({m^2_t\over \Lambda^2_{NP}} \right) \ \ , \\[0.1cm]
|f_{tB\Phi}| & \simeq &  {61.6\over\sqrt{1+645\, {m^2_t\over 
\Lambda^2_{NP}}}}~
\left ({m^2_t\over \Lambda^2_{NP}} \right) \ \ , \\[0.1cm]
|f_{tG\Phi}| & \simeq &
{m^2_t \sqrt{\pi}\over v \lam \sqrt{1+\frac{2}{3}\alpha_s}} \ \ .
\eqa
\par

At present the most important constraints on these couplings 
arise from the $Z$-peak experiments at LEP1/SLC, \cite{Zexp,
HagMS}.
In a near future the process $e^+e^-\to W^+W^-$ at LEP2, 
is expected to give direct constraints on the bosonic operators
in (17-19) \cite{LEP2TGC, BMT}. In addition, if the Higgs boson is light
enough the processes
$e^+e^-\to HZ$ and $e^+e^-\to H\gamma$ will also produce constraints on
the 3 other bosonic operators  (20-22) \cite{HZ}. In Section 6, 
all these constraints will be presented together with 
the observability limits that could be derived from the
$e^-e^+\to t\bar t$ and $t \to b W$ processes.\par

\section{The  $e^- e^+ \to t \bar t $ amplitude}
\hspace{0.5cm}
As it has been mentioned in the Introduction, 
box diagrams are never important for   
calculating the leading NP effects in  
$e^- e^+ \to t \bar t $. The amplitude has therefore 
a tree-level structure with $\gamma$ and $Z$ exchange in the
t-channel. Therefore,  we only need to determine the
$Vt\bar t$, ($V=\gamma, Z$) vertex, whose most general
CP-conserving form is 
\bq
\label{eq:dgz}
-i \epsilon_\mu^V J^{\mu}_V = -i e_V \epsilon_\mu^V 
\bar u_t(p)[\gamma^{\mu}d^V_1(q^2)+\gamma^{\mu}\gamma^5d^V_2(q^2)
+(p-p')^{\mu}d^V_3(q^2)] v_{\bar t}(p') \ ,
\eq
where $\epsilon_\mu^V$ is the polarization of the vector boson
V. The outgoing momenta $(p,~p')$ refer to $(t,~\bar t) $ 
respectively and ~satisfy $q\equiv p+p'$.
The normalizations are determined by 
$e_{\gamma}\equiv e$ and $e_Z\equiv
e/(2s_Wc_W)$. 
The couplings $d^V_i$ are in general  $q^2$ dependent
form factors. The contributions to these 
\underline{from SM at tree level} are
\bq
d^{\gamma, SM0}_1 = {2\over3} \ \ , \ \
d^{Z, SM0}_1=g_{Vt}={1\over2}-{4\over3}s^2_W\ \ , \ \
d^{Z, SM0}_2=-g_{At}=-\, {1\over2} \ \ .
\eq
In addition, there exist SM contributions to these couplings 
at the 1-loop level, $d^{V, SM1}_i$,
whose leading large $m_t$ part is  
computed in Sect.5. Finally in Section 6 we calculate the leading
NP contributions to $d^V_i$. For the operators
($\O_{t2}$, $\O_{Dt}$, $\O_{tW\Phi}$, $\O_{tB\Phi}$), these 
arise at the tree-level. For 
$(\O_{qt}$, $\O^{(8)}_{qt}$, $\O_{tt}$, $\O_{tb}$,
$\O_{tG\Phi})$ and the six purely bosonic operators
$(\O_W~,~\O_{W\Phi}~,~\O_{B\Phi}~,~\O_{UW}~,~\O_{UB}~,~
\O_{\Phi2})$ we need to go to the 1-loop in order to
find a non-vanishing  leading-log contribution which is also
enhanced by a power of $\mtd$. 
Finally, for the operators $\O^{(8)}_{tb}~,
~\O_{qq}~,~\O^{(8)}_{qq}~,~\O_{t1}~,~\O_{t3}$ we get no
such leading NP contribution,  up to the 1-loop order.\par

The $e^-e^+\to t\bar t$ helicity amplitude is written as 
$F_{\lambda,\tau,\tau'}$, where
$\lambda\equiv\lambda(e^-)=-\lambda'(e+)=\pm 1/2$
denote the $e^-$, $e^+$ helicities, while
$\tau$ and $\tau'$ represent respectively 
the $t$ and $\bar t$ helicities. 
For completeness we also mention  that the
$(e^-,~e^+)$ incoming momenta are denoted as $(k,~k')$, while
the $(t,~\bar t)$ outgoing momenta are $(p,~p')$.
Using the couplings defined in (\ref{eq:dgz}),  we write
\bqa
\label{eq:fltt}
F_{\lambda,\tau,\tau'}&=&\sum_{V=\gamma,Z} 2\lambda e^2\sqrt{s}
(A_V-2\lambda B_V)\Bigg \{ d^V_1 [2m_t \sin\theta\delta_{\tau \tau'}
+\sqrt{s} \cos\theta(\tau'-\tau)
-2\lambda\sqrt{s}\delta_{\tau,-\tau'}]\nonumber\\
&&-d^V_2 2|\pvec| [\cos\theta\delta_{\tau,-\tau'}+2\lambda(\tau-\tau')]
-d^V_3 4|\pvec|^2 \sin\theta\delta_{\tau \tau'}\Bigg \} \ \ , \
\eqa
\noindent
with $A_{\gamma} =-{1/ s}$, $A_Z=g_{Ve}/( 4s^2_Wc^2_W D_Z)$,
$B_{\gamma} = 0$,
$B_Z=g_{Ae}/( 4s^2_Wc^2_W D_Z)$ and $g_{Ve}=-1/2+ 2s^2_W$,
$g_{Ae}=-1/2$, $D_Z=s-M^2_Z+iM_Z\Gamma_Z$. In (\ref{eq:fltt}), 
$\theta$ is the $(e^-,t)$ scattering angle in
the $(e^-,e^+)$ c.m. frame. The amplitude is
normalized so that the unpolarized $e^-e^+\to t\bar t$ 
differential cross section is given by
\bq
{d\sigma(e^-e^+\to t\bar t)\over ~d\cos\theta}=
{3 \beta_t\over128\pi s}\sum_{\lambda, \tau,
\tau'} |F_{\lambda, \tau, \tau'}|^2 \ \ ,
\eq
where $\beta_t=(1-{4m^2_t\over s})^{1/2}$ and the colour factor
has been included. We note that CP
invariance implies \cite{Chang}
\bq
F_{\lambda,\tau,\tau'}~ = ~F_{\lambda, -\tau', -\tau}\ \ ,
\eq
which is of course satisfied by (\ref{eq:fltt}). In fact, at the
level of approximations used in constructing (\ref{eq:fltt}),
CPT implies also
\bq
F_{\lambda,\tau,\tau'}~ = ~F_{\lambda, -\tau', -\tau}^* \ \ ,
\eq
which indicates that all helicity amplitudes must be real. \par

Because of CP and CPT symmetries, it turns out that the simple
$t$ (or $\bar t$) decay distribution contains all information that
can be extracted from the amplitudes in (\ref{eq:fltt}). Thus,
 nothing more can be learned by considering the combined decay
distributions of $t$ and $\bar t$ simultaneously. It is,
therefore, sufficient to consider only the simple spin
density matrix of the produced $t$ or $\bar t$. For  top-quark , 
this is  
\bq
\rho^{L,R}_{\tau_1 \tau_2}=\sum_{ \tau'}
F_{\lambda, \tau_1, \tau'}F^*_{\lambda, \tau_2, \tau'}
 \ \ ,
\eq
where $L,R$ correspond respectively to 
$\lambda\equiv\lambda(e^-)=-\lambda'(e+)=\mp 1/2$.
There are only six independent elements 
(real in our case), $\rho^{L,R}_{++},
\rho^{L,R}_{--}$ and $\rho^{L,R}_{+-}=\rho^{L,R}_{-+}$, which can be 
measured through the
top production and  decay distributions. 
 Each $\rho$ element
has a typical angular distribution given in terms of the form
 $(1+\cos^2\theta)$,
$\sin^2\theta$ and $\cos\theta$, producing symmetrical and
asymmetrical $\theta$-distributions. Combining this  with the decay
angular information, various $\rho$ elements can be isolated.
Performing such measurements at a few $e^-e^+$ energies, even
for unpolarized beams, one can define a sufficient number of 
independent quantities that can be used to determine the six couplings
$d^V_i$, $V=\gamma, Z$, $i=1,3$. 
Many of these quantities, (and in particular those of interest
here), are forward-backward asymmetries in the angular
distribution of physical observables concerning the produced
top-quark. QCD effects to these observables are probably 
small \cite{Lampe, Kuhn}, and in any case they should be
incorporated in our formalism in the future.\par

Electron beam
polarization should provide an independent and maybe cleaner
way to disentangle these couplings,
through the separation
of left-handed ($ L \Leftrightarrow \lambda=-1/2 $) and right-handed
($ R \Leftrightarrow  \lambda=+1/2 $) contributions. Thus, in
addition to the unpolarized quantities, we  would then also have
the $L-R$ ones. This way, the information from the 
usual unpolarized
$L+R$ integrated cross section $\sigma(e^-e^+ \to t \bar t)$, 
will be augmented by the availability of also $\sigma^L$ and 
$\sigma^R$ allowing us to measure the 
integrated left-right asymmetry $A_{LR}$ defined as
$(\sigma^L-\sigma^R)/(\sigma^L+\sigma^R)$. 
Similarly, any forward-backward asymmetry constructed for 
the unpolarized ($L+R$) case will be ~accompanied  by 
the corresponding one in the polarized ($L-R$)
case. Details are given in Appendix B, where we thus
define the six forward-backward asymmetries $A_{FB}$, ~$A_{FB,pol}$,
~$H_{FB}$, ~$H_{FB,pol}$, ~$T_{FB}$, ~$T_{FB,pol}$.\par

\section{$t\to Wb$ decay amplitudes and induced asymmetries.}
\hspace{0.5cm}
The $t(p_t) \to W^+(p_W) b(p_b) $ decay,
where $(p_t,~p_W,~p_b)$ are the related momenta, 
will be used to construct the asymmetries mentioned in the last
paragraph of the previous section which 
are sensitive to the NP couplings affecting the $e^-e^+\to t \bar
t$, and to (a lesser extent) the ones  determining $ t \to bW$.
To describe the NP effects in the $t$ decay, we write  
the general $t \to W^+ b $ vertex in terms of four 
invariant couplings, related
through CPT invariance to the other four invariant couplings 
for $\bar t\to W^-\bar
b$. These couplings are given  by
\bq
\label{eq:dw}
-i \epsilon^{W*}_\mu J_W^{\mu}=-i\, 
{g V^*_{tb}\over2\sqrt2}\epsilon^{W*}_\mu \bar
u_b(p_b)[\gamma^{\mu}d^W_1+\gamma^{\mu}\gamma^5d^W_2+(p_t+p_b)^{\mu}d^W_3+
(p_t+p_b)^{\mu}\gamma^5d^W_4]u_t(p_t)\ ,
\eq
where $\epsilon^{W*}_\mu$ is the $W$ polarization vector and
$V^*_{tb}$ is the appropriate CKM matrix element.
The couplings $d^W_i$ receive contributions from SM and NP.
The \underline{tree level SM contribution} is
\bq
d^{W, SM0}_1=-d^{W, SM0}_2=1 \ \ , \ \ d^{W, SM0}_3=d^{W, SM0}_4=0 \ \ 
\eq
The 
\underline{1-loop $m^2_t$-enhanced SM contribution} $d^{W, SM1}_i$
are computed in Sect. 5
and they also satisfy the relations
\bq
d^{W, SM}_1=-d^{W, SM}_2 \ \ , \ \ d^{W, SM}_3=d^{W, SM}_4
\eq
\par 

Finally the NP contributions to top decay are also given
in Appendix B and collected in Sect. 6. Here we only note that
the operators $(\O_{t3}~,~\O_{Dt}~,~\O_{tW\Phi} )$ contribute
already at the tree level, while
$\O_{qq}~,~\O^{(8)}_{qq}~,~\O_{tG\Phi}$, 
$\O_{W\Phi}~,~\O_{B\Phi}~,~\O_{UW}~,~\O_{\Phi2}$ ~supply 
1-loop "leading-log"  
contributions, enhanced by powers of $m^2_t$. The remaining $\O_{qt}$,
$\O^{(8)}_{qt}$, $\O_{tt}$, $\O_{tb}$, $\O^{(8)}_{tb}$, $\O_{t1}$,
$\O_{t2}$, $\O_{tB\Phi}$, $\O_W$ and $\O_{UB}$ give no such
 contribution to $t\to bW$, up to this order.\par

In Appendix B we give the explicit forms of the 
asymmetry observables,
sensitive to the production couplings defined in (\ref{eq:dgz}),
and the decay ones in (\ref{eq:dw}). Since we expect the NP
effects to be small, we are only interested in observables
that are sensitive to the interferences between the NP and the 
tree-level SM effects.
Below, we first comment on
the asymmetries sensitive to the decay couplings and 
then on the production ones.\par

The $t \to b W^+ \to b l^+ \nu $
observables  which are interesting to measure are those 
sensitive to the interference between the NP and the tree-level 
SM effects. It turns out that these depend only on 
the combinations $(d^{W, NP}_1-d^{W, NP}_2)$ and 
$(d^{W, NP}_3+d^{W, NP}_4)$. The  
 $d^W_{3,4}$ couplings  have the
peculiarity of leading only to a W, longitudinally polarized
along the t 
quark momentum, whereas $d^W_{1,2}$ contribute to both the
transverse and 
longitudinal W's. Because of this, only the
combination $(d^{W, NP}_3+d^{W, NP}_4)$ can contribute linearly
to the asymmetries relevant for the top decay distribution. 
In Appendix B, two versions of such an asymmetry are given,
referring to the angular distribution of the lepton coming from 
the semileptonic $t$ or $\bar t$ decay. \par

The other combination $(d^{W, NP}_1-d^{W, NP}_2)$ cannot be seen
at linear order through asymmetries. 
For a physical quantity sensitive to it, 
we have to look at the partial width  
$\Gamma(t \to Wb)$  
\bqa
&&\Gamma(t\to Wb)={G_F|V^*_{tb}|^2 (m^2_t-M^2_W)^2\over 16 \pi \sqrt{2}
\, m^3_t}
\Big \{[(d^W_1)^2+(d^W_2)^2] (m^2_t+2M^2_W) \nonumber\\
&&+[(d^W_3)^2+(d^W_4)^2] (m^2_t-M^2_W)^2
+ 2 [d^W_3 d^W_1 - d^W_4 d^W_2)] m_t(m^2_t-M^2_W) \Big \} \  
, \eqa
where it appears multiplied by the CKM matrix element also.
Unfortunately, the width $\Gamma(t\to Wb)$  
cannot be directly measured to the necessary accuracy. 
Only indirectly can this be done, either using the fusion
process $W\gamma \to tb$ accessible at an $e^-e^+$ collider
in the $e\gamma$ mode (through laser backscattering);
or using reactions like $ q\bar q\prime \to t \bar b$ and 
$Wg\to t\bar b$ accessible at the Tevatron and 
LHC colliders respectively. For such a
measurement, one can expect
an accuracy of only about 20 to 30 \% for  
$(d^{W, NP}_1-d^{W, NP}_2)$ \cite{fusion, Heinson}.\par

We next turn to the asymmetries sensitive to the
production couplings defined in
(\ref{eq:dgz}). For constructing them,  we
need a description of the  angular
distribution of top production and decay. 
Below, we only give
expressions for completely longitudinally polarized
beams. In such a case, the angular distribution for 
$e^-e^+ \to \bar t t(t\to bW \to b  l \nu_l))$ can 
be written as
\bq
\frac{d\sigma^{L,R}}{d\cos \theta}={ 3 \beta_t\over32\pi
s} ~\sum_{\tau_1 \tau_2 \tau'}
F_{\lambda \tau_1 \tau' }F^*_{\lambda \tau_2 \tau'}
t_{\tau_1\tau_2} \ \ , \
\eq
where $(L,R)$ correspond to 
$\lambda\equiv\lambda(e^-)=-\lambda'(e+)=\mp 1/2$ respectively
and  $t_{\tau_1\tau_2}$ is the $t$-quark decay matrix
 constructed in terms of the helicity amplitude $\M_\tau(t \to b W^+
\to b l^+ \nu)$, with $\tau$ being the top helicity.  
We thus have 
\bq
t_{\tau_1\tau_2}={(2 \pi)^4 \over 2m_t\Gamma_t}
\sum_{spins}\M_{\tau_1}\M^*_{\tau_2} d \Phi_3(bl\nu_l)
\ \ ,
\eq 
where $\Gamma_t$ is the total top-width, 
$\sum_{spins}$ means summation over the final $(b,l^+,\nu_l)$ spins
and $d\Phi_3(bl\nu_l)$ is the usual 3-body phase space
describing $t$ decay in its rest frame,  \cite{PDG}. 
 
It is convenient to express the 3-body phase space in terms of the
Euler angles determining the $t$-decay plane. 
We start from the process $e^-(k) e^+(k') \to
t (p) \bar t(p') $ in the center of mass frame,
where with $\theta$ we denote the $(e^-,t)$ scattering angle. 
The $t$-quark rest frame (called t-frame hereafter)
is then defined with its z-axis
along  the top-quark momentum; the 
x-axis is taken in the $(t~\bar t)$ production plane, so that
the y-axis is perpendicular to 
it and along the
direction of   $\kvec \times \pvec$ of the
$e^-$ and $t$ momenta. 
In this $t$-frame  we define the top decay plane through
the Euler angles $(\varphi_1, \vartheta_1, \psi_1)$
described in Appendix B. In addition, within the top 
decay plane we define $\theta_l$ as the 
angle between the lepton momentum and the top momentum, after
having boosted to the W-rest
frame \cite{Lampe}. It is related to the 
$l^+$ energy in the $t$-frame by
\bq
E_l=|\overrightarrow p_l| =~\frac{\mtd +\mwd -\cos\theta_l
(\mtd-\mwd)}{4 \mt} \ \ ,
\eq
where the $(b,l^+)$ masses are neglected.
In terms of the Euler angles, the 3-body phase space becomes  \cite{PDG}
\bq
\delta((p_l+p_\nu)^2-\mwd) 
d\Phi_3(bl\nu_l) \Rightarrow  
{(m^2_t-M^2_W)\over64m^2_t(2\pi)^9}d\varphi_1d\cos
\vartheta_1d\psi_1 d\cos\theta_l \ \ ,
\eq\par
\noindent
after including the constraint that the $l\nu$ pair
lies at the $W$ mass shell. Like $\rho$, the matrix $t_{\tau_1\tau_2}$
also involves three real independent elements $(\tau_1\tau_2)=(++)$,
$(--)$ and $(+-)$. They
are explicitly written  in terms of the above angles
in Appendix B. Using these, we construct the three  forward-backward
asymmetries for the unpolarized case $(L+R)$ and another three
for the polarized one $(L-R)$, which
can be used to measure the production couplings in 
(\ref{eq:dgz}).\par

\section{Leading $m_t$-enhanced SM contributions at 1-loop.}
\hspace{0.5cm}
In the present section we study the 1-loop, $\mt$-enhanced SM
contributions to $e^-e^+\to  t \bar t$ and $t\to bW$.
Technically this means that we consider the large $\mt$ limit
of the 1-loop diagrams, 
keeping $m_H/\mt$ and $s/\mtd\equiv q^2/\mtd $ finite.
Such a study is useful for checking the possible appearance of
any large $m_t$ effect. It is also 
instructive for comparison with the corresponding  NP effects.
The relevant diagrams supplying such $\mtd$ enhancements
consist of triangular vertex-diagrams 
for the $\gamma t\bar t$ ,$Zt\bar t$, 
$Wtb$ vertices, and also 
of  the $t$ and $b$ self-energy diagrams;
involving exchanges of goldstone bosons
and the physical Higgs. Diagrams involving gauge boson
exchanges, or box diagrams, cannot
generate $m^2_t$-enhancements. We have checked that these
contributions to the form factors in (\ref{eq:dgz},
\ref{eq:dw}) are gauge invariant and determine  
the complete SM large-$q^2$, large-$\mt$ effect
in  $e^-e^+\to t\bar t$ and $t\to bW$. 
Note that we leave aside the gauge boson 
($\gamma, Z, W$) self-energy
contributions which are universal (i.e. not related to the $t\bar t$
channel) and are taken into account in the usual renormalization
procedure, \cite{Zcor}. 
For $q^2 \gsim 4 \mtd $, the resulting 1-loop contributions to 
the form factors defined in 
(44) is given by
\bqa
d^{\gamma, SM1}_1(q^2) &=&
-C\Big [{8+I_{se}\over3}+{1\over3}I_0-{1\over6}J_1+{1\over3}(I_2
+I_{2H})-{2\over3}J_{4H}\Big ] \ , \\  
d^{\gamma, SM1}_2(q^2) &= & 
-C \Big [{1\over3}+{1\over3}I_0+{1\over6}J_1\Big ] \ ,\\
d^{\gamma, SM1}_3(q^2)& = &
{2 C\over 3 m_t}\Big  [{1\over2}J_{2H}-J_0 \Big ] \ , \\
d^{Z,SM1}_1(q^2)&=&
C \Big [- ~ {3+I_{se}\over4}+{2s^2_W\over3}(8+I_{se})
+{2s^2_W\over3}I_0\nonumber\\
&&+ ~{1\over2}(1-{2s^2_W\over3})J_1
-{1\over4}(1-{8s^2_W\over3})(I_2+I_{2H}
-2J_{4H}) \Big ] \ , \\
d^{Z,SM1}_2(q^2)&=&
C \Big [{3+I_{se}\over4}+{2s^2_W\over3}+{2s^2_W\over3}I_0
-{1\over2}(1-{2s^2_W\over3})J_1+I_{1H}\nonumber\\
&&-~{1\over4}(I_2+I_{2H})+{1\over2}J_{3H}]\ , \\
d^{Z,SM1}_3(q^2) &=&
{C\over m_t}~\Big [{1\over4}(1-{8s^2_W\over3})J_{2H}
-(1-{4s^2_W\over3})J_0 \Big ] \ \ ,
\eqa
with $C=g^2\mtd/(64 \pi^2\mwd)$, and 
\bqa
&&I_{se}=-2\int^1_0 dx (1-x) \ln[x^2+\zeta(1-x)]~=~3~- \nonumber \\
&& 2\zeta \left (1-\frac{\zeta}{4} \right )\ln (\zeta) - 
\zeta -(\zeta- 2)
\sqrt{\zeta(\zeta-4)-i\epsilon} \,
\mbox{Arcosh} \left (\frac{\sqrt {\zeta}}{2}-
i\epsilon \right ) \ ,
\eqa
\bq
I_0=2\int_0^1 dx_1\int_0^{1-x_1} dx_2
\ln[(x_1+x_2)(x_1+x_2-1)-4\eta x_1x_2] \ ,
\eq
\bq
I_{1H}=2\int\int dx_1dx_2 \ln[(x_1+x_2-1)^2-4\eta x_1x_2+\zeta
x_1]\ ,
\eq
\bq
I_{2}=2\int\int dx_1dx_2 \ln[(x_1+x_2)^2-4\eta x_1x_2]\ ,
\eq
\bq
I_{2H}=2\int\int dx_1dx_2 \ln [(x_1+x_2)^2-4\eta x_1x_2+
\zeta(1-x_1-x_2) ]\ ,
\eq
\bq
J_0=2\int\int dx_1dx_2 {(x_1+x_2)(x_1+x_2-1)\over
(x_1+x_2)(x_1+x_2-1)-4\eta x_1x_2} \ ,
\eq
\bq
J_1=2\int\int dx_1dx_2 {(x_1+x_2-1)\over
(x_1+x_2)(x_1+x_2-1)-4\eta x_1x_2} \ ,
\eq
\bq
J_{2H}=2\int\int dx_1dx_2 \Big [{(x_1+x_2)(x_1+x_2-2)\over
(x_1+x_2)^2-4\eta x_1x_2+\zeta(1-x_1-x_2)}
+{(x_1+x_2)^2\over(x_1+x_2)^2-4\eta x_1x_2}\Big ]\ ,
\eq
\bq
J_{3H}=2\int\int dx_1dx_2
{2(1-{\zeta\over4}(x_1+x_2-1))\over
(x_1+x_2)^2-4\eta x_1x_2+\zeta(1-x_1-x_2)}\ ,
\eq
\bq
J_{4H}=2\int\int dx_1dx_2 {2-{\zeta\over2}(x_1+x_2-1)\over
(x_1+x_2)^2-4\eta x_1x_2+\zeta(1-x_1-x_2)}\ ,
\eq
with $\eta=(q^2 +i \epsilon )/(4\mtd)$ and $\zeta=m^2_H/ \mtd $.
The double integration runs over the interval $[0,1]$ for $x_1$,
and $[0,1-x_1]$ for $x_2$.
These integrals ~have been computed partially analytically and 
partially
numerically\footnote{We thank N.D. Vlachos for his help in this
numerical computation.}.\par

It can be noted from these results that close to 
threshold (\ie~ $q^2 \sim
4m^2_t$), there is a rather strong ${m^2_t/m^2_H}$ dependence 
arising from the triangle diagrams involving physical Higgs
exchange. If we put $\mt \gg m_H$, then the SM contribution 
acquires infrared-type singularities, which are obviously related
to the stability of scalar sector requiring $\mt \sim m_H$ for 
physically acceptable $\mt$ masses, 
\cite{Higgsstab}.\par

In Fig.1 we plot (in units of the coefficient $C\simeq 0.003$),
these SM contributions to the six
form factors, 
as functions of $\sqrt{q^2} \equiv (2E_e)$,   for
$m_H=0.1~TeV$ and $q^2 > 4m^2_t$. For
$q^2\gsim (0.5~TeV)^2$ we find from this figure 
an effect of the order of 1\%
for the vector $\gamma t\bar t$ coupling, and of the order of 3
per mille for the other vector and axial couplings. The 
"derivative" $d^{\gamma, Z}_3$ couplings are somewhat weaker.\par

We next turn to the  $m^2_t$-enhanced SM corrections to the 
$t\to b W$ decay. 
The resulting 1-loop contributions, 
expressed in terms of the decay couplings defined in
(51), is
\bq 
d^{W, SM1}_1~=~ -d^{W,
SM1}_2=- ~ C\left [{5\over4}+{1\over4}I_{se}+{1\over2}H_1
\right ]\ \ , 
\eq
\bq
d^{W, SM1}_3=d^{W, SM1}_4=- ~ {C\over2m_t}\,
[1+H_2] \ \ ,
\eq
where
\bq
H_1=2\int\int dx_1dx_2 \ln [ x_1(x_1+x_2)-\zeta (x_1+x_2-1) ]
\ ,
\eq
\bq
H_2=2\int\int dx_1dx_2
{x_1(x_1+x_2-2)\over x_1(x_1+x_2)-\zeta (x_1+x_2-1)} \ .
\eq

\vspace{2cm}
\begin{center}
Fig.1: 1-loop SM contributions to the $d^\gamma_j$ and $d^Z_j$
form factors defined in (\ref{eq:dgz})), as functions of
$\sqrt{q^2}=2E_e$.\\
\end{center}
\vspace{2cm} 

For $m_H=0.1~TeV$ these equations give 
\bq
d^{W, SM1}_1 ~=~ -d^{W, SM1}_2~= ~ C (-0.92 ) \ ,
\eq
\bq
d^{W, SM1}_3 ~=~ d^{W, SM1}_4 ~= ~ C \left
(\frac{0.0728}{2m_t}\right ) \ .
\eq
Comparing the definition of the various couplings given in (51),
with the relations (75,76), we remark that the SM $\mt^2$-enhanced
couplings only affect the left-handed $b_L$ field. This is 
also obvious from the structure of the relevant diagrams.

 
\section{The NP effects.}
\hspace{0.5cm}
In Appendix A we enumerate the relevant diagrams giving the 
~leading NP
contribution to $e^-e^+ \to t \bar t$ and $t\to bW$, for each kind of
operator. Since box diagrams are never important,
the leading NP contribution to $e^-e^+ \to t \bar t$
 can be expressed in terms of the NP contributions to the
form factors introduced in (\ref{eq:dgz}). Here, 
$s=q^2\geq 4\mtd$ is understood. 
As already stated, the operators
$(\O_{t2}~,~\O_{Dt}~,~\O_{tW\Phi}~,~\O_{tB\Phi})$ 
give tree-level contributions, which are 
\bq
d^{\gamma,NP}_1(s)=-~{4\sqrt2  M_W\over \ e^2 \mt}(\swd f_{tW\Phi}
+\sw \cw f_{tB\Phi} )\ \ ,
\eq
\bq
d^{\gamma,NP}_3(s)={2\sqrt2 M_W\over e^2 \mtd} (\swd f_{tW\Phi}
+ s_W c_W f_{tB\Phi}) \ \ ,
\eq
\bq
d^{Z,NP}_1(s)=-~{2 \mwd\over g^2 \mtd }\, f_{t2}
+{8\sqrt2 M_W\over g^2 \mt}( - \cwd f_{tW\Phi} + \sw
\cw f_{tB\Phi} )\ \ , 
\eq
\bq
d^{Z,NP}_2(s)=-~{2 \mwd\over g^2 \mtd }\, f_{t2}\ \ ,
\eq
\bq
d^{Z,NP}_3(s)=-~{M_W\over\sqrt2 g \mtd }\, f_{Dt}
+{4\sqrt2 M_W\over g^2 \mtd}( \cwd f_{tW\Phi}
- \sw \cw f_{tB\Phi} )\ \ .
\eq \par

We next turn to the operators contributing only at the 
1-loop level. 
As explained above, we consider only the leading NP effect 
determined by the divergent part\footnote{We always use
dimensional regularization.} of the Feynman integrals,
provided it is enhanced by some power of $\mtd$. 
The contributions from the "top"-involving operators are
then expressed in terms of 
\bq
\label{eq:Fi}
F_i\equiv ~{1\over16\pi^2}\ln \left (\Lambda^2\over\mu^2
\right ) {f_i\over m^2_t} \ \ , 
\eq
with $\Lambda$ being the divergent integral ~cutoff identified 
with the NP scale $\lam$, and $\mu \sim \sqrt{s}$
being the scale where the effective coupling is measured.
For the purely bosonic operators, due to their
different normalization implied by (24,25), we
should replace in (\ref{eq:Fi}) $f_i/\mtd \to (\lw/\mwd,~
f_W/\mwd,~f_B/\mwd,~
d/v^2,~d_B/v^2,~f_{\Phi2}/v^2)$ for
($\O_W$, $\O_{W\Phi}$, $\O_{B\Phi}$,
$\O_{UW}$, $\O_{UB}$, $\O_{\Phi2}$) respectively.  
We thus get
\bqa
&&d^{\gamma,NP}_1(s)=~{s\over6} F_{qt}+ {8 s \over9 } F_{qt(8)}
-{8s\over9} F_{tt}+{s\over3}F_{tb}
+~{256\sqrt2 g_s m_t M_W\over 9g}F_{tG\Phi} \nonumber\\
&& +~ {g^2 s\over4} F_W - ~ 
{g^2 m^2_t s\over 32 M^2_W}[F_{W\Phi}+F_{B\Phi}]
-2m^2_t F_{UW}-{10\mtd \over3} F_{UB} \ \ ,
\eqa
\bq
d^{\gamma,NP}_2(s)=-\,{s\over18}F_{qt}- {8 s\over 27} F_{qt(8)}
-{8s\over9}F_{tt}+{s\over3}F_{tb}
-{g^2 s\over4} F_W+{g^2 m^2_t s\over 32M^2_W}
[F_{W\Phi}-3F_{B\Phi}]\ ,
\eq
\bq
d^{\gamma,NP}_3(s)=-\, {128\sqrt2 g_s M_W\over9g}F_{tG\Phi}+m_t F_{UW}
+{5 \mt \over3} F_{UB} \ ,
\eq
\bqa
&&d^{Z,NP}_1(s)=-~{s s^2_W\over3}F_{qt}-~ {16 s s^2_W\over 9}F_{qt(8)}
-4\left (m^2_t-{4s s^2_W  \over9}\right )F_{tt}-{2s s^2_W\over3}
F_{tb} \nonumber\\
&&+~{64 \sqrt2 g_s(3-8s^2_W) m_t M_W\over 9g}F_{tG\Phi}
+{ s c^2_W g^2\over2} F_W -{g^2 m^2_t s\over16M^2_W}
[c^2_W F_{W\Phi}-s^2_W F_{B\Phi}]\nonumber\\
&& -4m^2_t c^2_W F_{UW}
+~ {20m^2_t s^2_W \over3} F_{UB} \ \ ,
\eqa
\bqa
&&d^{Z,NP}_2(s)=-(m^2_t-{s s^2_W\over 9})F_{qt}-
\left ({16\over3}\right )(m^2_t-{s s^2_W\over 9})F_{qt(8)}
-4(m^2_t-{4s s^2_W \over9})F_{tt}\nonumber\\
&&-~{2s s^2_W\over3} F_{tb}
-{s c^2_W g^2\over2} F_W +{g^2 m^2_t s\over16M^2_W}
[c^2_W F_{W\Phi}+3s^2_W F_{B\Phi}] -4m^2_t F_{\Phi2}
\ , 
\eqa
\bq
d^{Z}_3(s)=-~{32\sqrt2 g_s(3-8s^2_W) M_W\over9 g}F_{tG\Phi}
+2m_t c^2_W F_{UW}-{10 m_t s^2_W \over3} F_{UB}\ .
\eq \par

Similarly, for the $t\to b W$ couplings defined in
(\ref{eq:dw}), the non vanishing tree level NP contributions  
are
\bq
d^{W,NP}_1={2 \mwd \over g^2 \mtd}f_{t3} -~{4\sqrt2  M_W\over
g^2\mt } f_{tW\Phi} \ \ ,
\eq
\bq
d^{W,NP}_2={2 \mwd \over g^2 \mtd}f_{t3} +~{4\sqrt2  M_W\over
g^2\mt } f_{tW\Phi} \ \ ,
\eq
\bq
d^{W,NP}_3={4\sqrt2 M_W\over
g^2 m^2_t} f_{tW\Phi} -~\frac{\mw}{g\sqrt{2} \mtd} f_{Dt} \ 
\ ,
\eq
\bq
d^{W,NP}_4={4\sqrt2 M_W\over
g^2 m^2_t} f_{tW\Phi} -~\frac{\mw}{g\sqrt{2} \mtd} f_{Dt} \ 
\ .
\eq
Correspondingly, the non vanishing 1-loop $m^2_t$-enhanced NP 
contributions (for the operators which do not have any
tree-level ones) are
\bqa
&&d^{W,NP}_1={m^2_t\over2}F_{qq}+{8 \mtd\over3}F_{qq(8)}
+{32\sqrt2 m_t M_W g_s\over3g}F_{tG\Phi}
+{g^2 m^2_t(13s^2_W-3)\over48c^2_W}F_{W\Phi}\nonumber\\
&&+~{g^2 m^2_t s^2_W\over48c^2_W}F_{B\Phi}-2m^2_tF_{UW}
+2m^2_tF_{\Phi2}\ \ ,
\eqa
\bqa
&&d^{W,NP}_2={m^2_t\over2}F_{qq}+{8 \mtd \over3} F_{qq(8)}
-{32\sqrt2 m_t M_W g_s\over3g}F_{tG\Phi}
-{g^2 m^2_t(13s^2_W-3)\over48c^2_W}F_{W\Phi}\nonumber\\
&&-~{g^2 m^2_t s^2_W\over48c^2_W}F_{B\Phi}+2m^2_tF_{UW}
-2m^2_tF_{\Phi2}\ \ ,
\eqa
\bqa
&&d^{W,NP}_3=-~{m_t\over2}F_{qq}-{8 \mt \over 3} F_{qq(8)}
-{32\sqrt2 M_W g_s\over3g}F_{tG\Phi}\nonumber\\
&&-~{g^2 m_t(11s^2_W-6)\over24c^2_W}F_{W\Phi}
-{5g^2 m_t s^2_W\over24c^2_W}F_{B\Phi}+2m_tF_{UW}
\ \ ,
\eqa 
\bqa
&&d^{W,NP}_4= {m_t\over2}F_{qq} +{8 \mt \over 3} F_{qq(8)}
-{32\sqrt2 M_W g_s\over3g}F_{tG\Phi}\nonumber\\
&&-~{g^2 m_t(11s^2_W-6)\over24c^2_W}F_{W\Phi}
-{5g^2 m_t s^2_W\over24c^2_W}F_{B\Phi}+2m_tF_{UW}
\ \ .
\eqa\par

\newpage 
We now review the NP effects of each operator on the various observables.
First note that the operators $\O^{(8)}_{tb}$ and $\O_{t1}$
give no contribution (within our approximations) to 
either production or decay. The effects of the rest
are illustrated in Table 1, where we give the NP 
couplings used and the implications for the $t\to bW$ decay 
observables. The effects on top production through $e^-e^+ \to t
\bar t$  are presented in\footnote{Note that in the 
figures, the signatures of $\O^{(8)}_{qt}$ are never shown 
~explicitly,
since they are identical to those from $\O_{qt}$, apart from an
overall normalization factor of 16/3.} Figs.2-9c. \par

\vspace{1.2cm}
\begin{center}
Fig.2: NP effects for $\O_{qt}$ as seen in the unpolarized 
cross section $\sigma(e^-e^+ \to t \bar t)$, as a function of 
the c.m. energy $2E_e$. The indicated value of the $f_{qt}$ coupling 
is chosen to create a 30\% effect.
Similar results are obtained for the contributing 
operators $\O_{t2}$, $\O_{Dt}$, 
$\O_{tW\Phi}$, $\O_{tB\Phi}$, $\O^{(8)}_{qt}$, $\O_{tt}$, $\O_{tb}$,
$\O_{tG\Phi}$ and $\O_W$, $\O_{W\Phi}$, $\O_{B\Phi}$, $\O_{UW}$,
$\O_{UB}$ and $\O_{\Phi2}$, using the NP couplings shown in the 
second column of Table 1.\\

\end{center}
\newpage

\vspace{2cm}
\begin{center}
Fig.3: NP effects  on the
unpolarized forward-backward asymmetry in the 
~differential cross section $A_{FB}(e^-e^+ \to t \bar t)$,  
as function of the c.m. energy $2E_e$, for the NP couplings in
Table 1.  
~~(a): 4-quark NP operators; $\O_{tt}$ effect is similar to $\O_{tb}$.
~~(b):  2-quark NP operators;  $O_{tB\Phi}$ gives similar effects to
$\O_{tW\Phi}$. 
~~(c): purely bosonic NP operators; the effects of $\O_{W\Phi}$ 
and $\O_{UB}$ are similar to those from $\O_W$ and $\O_{UW}$
respectively. \\
\end{center}
\newpage

\vspace{2cm}
\begin{center}
Fig.4:  NP effects on the
~unpolarized forward-backward asymmetry 
in the top average helicity $H_{FB}(e^-e^+ \to t \bar t)$, 
as function of the c.m. energy $2E_e$, for the NP couplings in
Table 1.
~~(a): 4-quark NP operators.
~~(b): 2-quark NP operators; the $\O_{tB\Phi}$ effect is similar 
to $1/3$ the one from $\O_{tW\Phi}$; while the effect from 
$\O_{Dt}$ is very small. 
~~(c): purely bosonic operators; $\O_{W\Phi}$ behaves  
similarly to $\O_W$; while $\O_{UB}$ gives a very 
small contribution.\\
\end{center}

\vspace{0.5cm}
\begin{center}
Fig.5:  NP effects on the
~unpolarized forward-backward asymmetry 
in the top transverse polarization $T_{FB}(e^-e^+ \to t \bar t)$, 
as function of the c.m. energy $2E_e$, for the NP couplings in
Table 1. 
~~(a): 4-quark NP operators.
(b1),(b2):  2-quark NP operators.  
~~(c): purely bosonic operators; $\O_{W\Phi}$ behaves  
similarly to $\O_W$; $\O_{UW}$ gives very small effect for
$d=-5.94$, while for $d=5.94$ the effect is similar to the
one from $\O_{\Phi2}$ for $f_{\Phi2}=11.9$. \\
\end{center}
\vspace{2cm}
\begin{center}
Fig.6: NP effects, for longitudinally polarized beams, 
on the Left-Right asymmetry $A_{LR}$, for 
the NP couplings in Table 1. 
~~(a): 4-quark NP operators; $\O_{tt}$ effect is similar to $\O_{tb}$. 
~~(b): 2-quark NP operators; $\O_{tG\phi}$ effect is very small. 
~~(c): purely bosonic NP operators; the effect of $\O_{W\Phi}$
is of similar magnitude but opposite sign to the one from $\O_W$;
while the $\O_{\Phi 2}$ effect is very small.\\
\end{center}
\vspace{2cm}
\begin{center}
Fig.7: NP effects on the
polarized forward-backward asymmetry in the 
~differential cross section $A_{FB, pol}(e^-e^+ \to t \bar t)$,  
as function of the c.m. energy $2E_e$, for the NP couplings in
Table 1. 
~~(a): 4-quark NP operators; $\O_{tt}$ effect is similar to $\O_{tb}$.
~~(b):  2-quark NP operators; $\O_{t2}$ effect is similar to 
$\O_{tG\Phi}$ give. 
~~(c): purely bosonic NP operators; $\O_{W\Phi}$ and $\O_W$
effects are of equal magnitude but of opposite sign to the 
$\O_{B\Phi}$ effect. \\
\end{center}
\vspace{2cm}
\begin{center}
Fig.8:  NP effects on the
~polarized forward-backward asymmetry 
in the top average helicity $H_{FB, pol}(e^-e^+ \to t \bar t)$, 
as function of the c.m. energy $2E_e$, for the NP couplings in
Table 1.
~~(a): 4-quark NP operators. 
~~(b): 2-quark NP operators; $\O_{tB\Phi}$ and $\O_{tW\Phi}$
behave similarly; $\O_{Dt}$ effect is very small. 
~~(c): purely bosonic operators; $\O_{UB}$ behaves  
similarly to $\O_{UW}$; $\O_{W\Phi}$ effect is similar to $\O_W$.\\
\end{center}
\vspace{1cm}
\begin{center}
Fig.9:  NP effects on the
polarized forward-backward asymmetry 
in the top transverse polarization $T_{FB, pol}(e^-e^+ \to t \bar t)$, 
as function of the c.m. energy $2E_e$, for the NP couplings in
Table 1.
~~(a): 4-quark NP operators.
~~(b): 2-quark NP operators; The size of the $\O_{tG\Phi}$
effect $\sim (\frac{1}{2}~-~\frac{1}{3})\O_{t2} $; while
the effect from $\O_{tW\Phi}$ is a rough average of 
the $\O_{tB\Phi}$ and $\O_{Dt}$ ones.  
~~(c): purely bosonic operators; the effects of $\O_W$, $\O_{W\Phi}$\
and $\O_{B\Phi}$ are not shown since they are small.\

\end{center}

\vspace{1cm}

In order to make the production effects clearly visible
in the figures, we have chosen the NP 
couplings such that the NP effect 
is about $\pm30\%$ of the SM prediction  on the
integrated cross section. A corresponding choice with respect to
the NP contribution to $\Gamma (t \to bW) $ has also been made
for those operators which contribute only to decay and not to 
production. The third column in Table 1 identifies the
operators giving non-vanishing contribution to top production
either at the tree or at the 1-loop level.
The rest of the columns in Table 1 describe the NP effects on  
$\Gamma( t \to b W)$ and the two forward-backward 
asymmetries constructed in Appendix B 
for the semileptonic top decay. \par

 \noindent
\begin{center}
\begin{tabular}{|c|c|c|c|c|c|} \hline
\multicolumn{6}{|c|}{Table 1: NP ~couplings and effects on production
and decay observables.} \\[.1cm] \hline
\multicolumn{1}{|c|}{$\O_i$} &
 \multicolumn{1}{|c|}{ $f_i,d,d_B,\lw $} &
  \multicolumn{1}{|c|}{$e^-e^+\to t \bar t$} &
    \multicolumn{1}{|c|}{$\Gamma(t\to bW)$} &
     \multicolumn{1}{|c|}{$D^1_{FB}$} &
      \multicolumn{1}{|c|}{$D^2_{FB}$} \\[0.1cm] \hline
  $SM$ &--& Yes & $1.5581$ & $0.2210$ & $-0.5384$  \\[0.1cm]\hline
  $+\O_{qt}$ & $(+,-)1.8$ & loop & No & No  & No  \\[0.1cm]\hline
  $+\O^{(8)}_{qt}$ & $(+,-) 0.34$ &loop & No & No & No  \\[0.1cm]\hline
  $+\O_{tt}$ & $(+,-) 0.27$& loop  & No & No   &No \\[0.1cm]\hline
  $+\O_{tb}$ & $(+,-)0.6$ & loop & No & No  & No   \\[0.1cm]\hline
$+\O^{(8)}_{tb}$& \null & No& No & No & No \\[0.1cm]\hline
$+\O_{qq}$ & $(+,-)3$ & No & 1.5583 & $0.2210$&$-0.5384$  \\[0.1cm]\hline
$+\O^{(8)}_{qq}$ & $(+,-)3$& No & 1.5644 & $0.2210$ & $-0.5384,-0.5383$
\\[0.1cm]\hline 
$+\O_{t1}$& \null & No & No & No & No \\[0.1cm]\hline 
$+\O_{t2}$ & $(+,-)0.12$ & tree & No & No   & No \\[0.1cm]\hline
$+\O_{t3}$ & $(+,-)0.9$ & No & $2.7910$ & $0.2211$ & $-0.5385,-0.5388$
\\[0.1cm]\hline
$+\O_{Dt}$ & $(+,-)0.51$ & tree &$0.9373,5.4581$ &$0.2211$
&$-0.5385$\\[0.1cm]\hline 
$+\O_{tW\Phi}$ & $(-,+)0.012$ & tree &$1.6619,1.4597$ &
$0.2407,0.2030$ & $-0.6720,-0.4491$ \\[0.1cm]\hline
$\O_{tB\Phi}$ & $(-,+)0.009$ & tree & No & No  & No \\[0.1cm]\hline 
$+\O_{tG\Phi}$ & $ (+,-)0.24$ &loop & $1.5041,1.6136$ &
$0.2117,0.2313$ & $-0.4860,-0.6034$ \\[0.1cm]\hline
$+\O_W$ & $(+,-)0.40 $ & loop & No & No & No  \\[0.1cm]\hline
$+\O_{W\Phi}$ & $(-,+)0.63 $ & loop & $1.5523,1.5639$ & $0.2219,0.2202$  &
$-0.5433,-0.5336$ \\[0.1cm]\hline 
$+\O_{B\Phi}$ & $ (-,+)0.75 $ &loop & $1.5600,1.5561$
&$0.2207,0.2214$ & $-0.5365,0.5404$ \\[0.1cm]\hline
$+\O_{UW}$ & $(-,+)5.94 $ &loop & $1.6770,1.4463$ &$0.2437,0.2022$ &
$-0.6964,-0.4388$ \\[0.1cm]\hline
$+\O_{UB}$ & $(-,+)5.35 $ & loop & No & No & No  \\[0.1cm]\hline
$+\O_{\Phi 2}$ & $(+,-)11.9$ &loop & $2.1235$ & $0.2210$ & $-0.5384$  \\
\hline
 \end{tabular}
\end{center}
\noindent

In Fig.2 we give the NP results for the integrated 
~unpolarized cross ~section $\sigma(e^-e^+ \to t \bar t) $
for the $\O_{qt}$ case. Similar
results would appear for  all other operators, if the associated
coupling constants take the values given in Table 1. 
The particular characteristics of each operator 
may then be studied by looking at the other observables appearing 
in the 4th-6th columns of Table 1 and in Figs.3a-9c.
In detail, for unpolarized beams, the forward-backward 
asymmetry $A_{FB}$ is illustrated in Figs.3a,b,c, 
the asymmetry $H_{FB}$ in Figs.4a,b,c, and
the asymmetry $T_{FB}$ in Figs.5a,b1,b2,c. 
Correspondingly, for ~longitudinally
polarized $e^\mp$ beams, the asymmetry $A_{LR}$
is illustrated in Figs.6a,b,c, while the forward-backward
asymmetry $A_{FB,pol}$ is in Figs.7a,b,c,
the asymmetry $H_{FB,pol}$ in Figs.8a,b,c 
and $T_{FB,pol}$ in Figs.9a,b,c.
Here, the "a"-figures refer to the 4-quark operators,
the "b"-figures to the 2-quark ones, and the "c" to the 
bosonic operators.
Occasionally in the figures, the results  for some  operators
or observables almost coincide, for the couplings chosen above.
Whenever this happens, it is just indicated in the figure caption.  \par

The values
for the coupling constants used in Table 1, 
are often unacceptably large, either because of the existing 
indirect experimental constraints, or because they would imply,
through unitarity, a very low NP scale. Nevertheless we used them in
order to make the NP effects in the figures clearly visible.

The expected luminosity for a future linear collider is 
commonly taken
as $80(s/TeV^2)$ $fb^{-1}$ per year.
Since, the SM cross section for $e^+e^-\to t\bar t$ is
about $170(TeV^2/s)~fb$, we expect a rate of
more than $10^4$ events/year, implying a statistical 
accuracy of $\sim 1\%$ for the various
physical quantities. 
The implied observability limits to various NP
couplings are presented in Table 2, where we also give
the present constraints \cite{Zexp}. In getting them,
we have always assumed that only one NP operator acts at a time.
Moreover, we have conservatively  assumed a total 
(statistical + systematical)
relative accuracy 
of 5\% on the integrated
cross section for unpolarized $e^\pm$ beams
and an absolute 5\% accuracy on the asymmetries
defined in Appendix B.  
More precise numbers require of course detailed Monte
Carlo analyses, taking into account the precise experimental 
conditions. Finally, using the unitarity
relations presented in Section 2, we translate the
observability limits on the various NP couplings, to
bounds on the highest related NP scales $\lam$, to which the
specific measurements are sensitive. These bounds are also
indicated in Table 2. In the last three columns we have indicated for
each operator separately, the constraints established in ref.\cite{bb}
and the expected LEP2 sensitivity.
In all cases except for the $Z \to b\bar b$ case, the absolute value of
the coupling constant is meant.
\par
The following comments should now be made concerning the
properties of the various operators and the observability limits
presented in Table 2:\\
\underline {Four-quark operators}\par

$\O_{qt}$ and $\O^{(8)}_{qt}$:\\
Both operators lead to the same effects (apart from an overall
normalization factor 16/3). They both contribute, at 1-loop,
only to the vector and axial form factors 
$d^{\gamma}_{1,2}$ and $d^{Z}_{1,2}$.   They do
not contribute to the top decay. So their modifications of the SM
predictions are always rather uniform as one can see on the figures.
The observability limits obtained either from the integrated cross
section or from
$H_{FB}$, appear to be just marginally
compatible with the LEP1 constraint from $ Z\to b\bar b $, to
which they also contribute at 1-loop, \cite{bb}.\par

$\O_{tt}$:  \\
It induces (through 1-loop) a purely right-handed NP effect
to the $\gamma t \bar t$ and $Zt\bar t$ couplings. 
There exist no $Z$-peak constraint on $\O_{tt}$. The
observability limit mainly comes from $H_{FB}$. No effect is
generated in top decay.\par

$\O_{tb}$:\\
Its effect is similar to the $\O_{tt}$ one.  However, it also
gives a purely right-handed contribution to $Z \to b \bar b$, 
which is thus providing
a LEP1 constraint. Its observability limit in Table 2, is
just allowed by the present  LEP1 results. No effect is
generated in top decay.\par

$\O^{(8)}_{tb}$:\\
This operator produces no effect in the processes studied here
or in $Z$-peak physics.\par

$\O_{qq}$  and $\O^{(8)}_{qq}$:\\
They contribute (at the 1-loop level) to the $t \to bW$ decay
couplings $d^W_j$ giving a $\sigma_{\mu\nu}$-type contribution
affecting only the right-handed $b_R$ field. They also give
a $\sigma_{\mu\nu}$ contribution to $Z \to b \bar b$. 
Both effects are too small to be observable for reasonable 
values of the coupling constants. 
They give no contribution
to the $\gamma t \bar t$ or $Z  t\bar t$ vertices.\par
 
\vspace{0.5cm}

\noindent
\begin{center}
\begin{tabular}{|c|c|c|c|c|c|c|} \hline
\multicolumn{7}{|c|}{Table 2: Sensitivity limits to "top" and
"bosonic"operators }\\[0.1cm]
\multicolumn{7}{|c|}{in terms of NP couplings and related NP
scales $\lam$ (TeV)} \\[.1cm] \hline
\multicolumn{1}{|c|}{Operator} &
  \multicolumn{1}{|c|}{$\sqrt{s}=0.5$ TeV} &
   \multicolumn{1}{|c|}{$\sqrt{s}=1$ TeV} &
     \multicolumn{1}{|c|}{$\sqrt{s}=2$ TeV} &
       \multicolumn{3}{|c|}{Other constraints from}\\[0.1cm]
\multicolumn{1}{|c|}{\null} &
  \multicolumn{1}{|c|}{($\lam$)} &
   \multicolumn{1}{|c|}{($\lam$) } &
     \multicolumn{1}{|c|}{($\lam$)} &
       \multicolumn{1}{|c|}{$\epsilon_i$}&  
       \multicolumn{1}{|c|}{ $Z \to b\bar b$} &
         \multicolumn{1}{|c|}{  LEP2 }	  \\[0.1cm] \hline
$\O_{qt}$ & 0.35(1.2)& 0.3(1.3) & 0.2(1.6) & --- &$-0.3\pm 0.2$ &--- \\[0.1cm]\hline
$\O^{(8)}_{qt}$ & 0.07(1.3) & 0.06(1.4) &0.04(1.8) &---& $-0.05\pm 0.03$
&--- \\[0.1cm]\hline
$\O_{tt}$ & 0.02(5.4) & 0.015(6.2) & 0.01(7.6) & --- &--- & --- 
\\[0.1cm]\hline
$\O_{tb}$ & 0.07(3.3) & 0.04(4.4) & 0.03(5.1) &---&$-0.3\pm 0.2$ &--- \\[0.1cm]\hline
$\O^{(8)}_{tb}$ & ----- & ----- & ----- & --- & --- & ---\\[0.1cm]\hline
$\O_{qq}$ & ----- & ----- & ----- &---& $38.\pm 22.$& --- \\[0.1cm]\hline
$\O^{(8)}_{qq}$ & --- & ---  & ---  &---& $8.\pm 5.$& --- \\[0.1cm]\hline
$\O_{t1}$ & ----- & ----- & ----- & ----- &---& --- \\[0.1cm]\hline
$\O_{t2}$ & 0.011(11.) & 0.016(9.1) & 0.017(8.9) & 0.01 &$0.3\pm 0.2$
&--- \\[0.1cm]\hline
\multicolumn{1}{|c|}{$\O_{t3}$} & 
\multicolumn{3}{|c|} {from decay 0.045 (6.5)} & 
\multicolumn{1}{|c|} {---} & 
\multicolumn{1}{|c|} {---} & \multicolumn{1}{|c|} {---} \\[0.1cm]\hline
$\O_{Dt}$ &0.036(2.6) & 0.03(2.9) &0.025(3.1) & 0.03 &$-0.12\pm 0.06$ &
--- \\[0.1cm]\hline
$\O_{tW\Phi}$ & 0.002(30.5) & 0.002(30.5) & 0.0015(35) &0.014 &
---& --- \\[0.1cm]\hline
$\O_{tB\Phi}$ & 0.0015(35.) & 0.0015(35.) &0.0013(38.) & 0.013
&---&---   \\ [0.1cm]\hline
$\O_{tG\Phi}$ & 0.02(10.) & 0.03(6.9) & 0.075(2.8) & ---&---&---
 \\[0.1cm]\hline
$\O_W$ & 0.05(1.6) & 0.04(1.7) & 0.02(2.5) &---&---& 0.1 \\[0.1cm]\hline
$\O_{W\Phi}$ & 0.08(1.6) & 0.06(1.8) & 0.04(2.2) &---&---& 0.1  \\[0.1cm]\hline
$\O_{B\Phi}$ & 0.025(5.0) & 0.02(5.6) & 0.01(7.9) &---&---& 0.1  \\[0.1cm]\hline
$\O_{UW}$ & 0.5(1.2) & 0.8(0.9) & 1.6(0.65) &---&---& 0.015  \\[0.1cm]\hline
$\O_{UB}$ & 0.5(1.6) & 0.6(1.45) & 1.2(1.0) &---&---& 0.05  \\[0.1cm]\hline
$\O_{\Phi 2}$ & 0.5(1.0) & 1.(0.7) & 2.4(0.5) &---&---& 0.01  \\ \hline
 \end{tabular}
\end{center}
\noindent

\underline {Two-quark operators.}\par

$\O_{t1}$: \\
No effect is generated in top production or decay or
in $Z$-peak physics.\par
 
$\O_{t2}$:\\
It gives a purely right-handed tree level contribution 
to the $Zt\bar t$ coupling. At the 1-loop level, it also
contributes to  
$\epsilon_1$ and to the $Zb \bar b$ vertex in a purely left-handed
way. Present constraints from $\epsilon_1$  
are marginal. There is no effect on the $\gamma t \bar t$ and $tbW$
vertices at tree level.\par
 
$\O_{t3}$:\\
At tree level, it produces a right-handed contribution to the
$tbW$ vertex. Its most important constraint should come from
$\Gamma (t \to b W)$. 
There exists no constraint from $Z$ peak physics.\par
 
$\O_{D t}$: \\
At the tree level, it contributes a derivative 
coupling to the $Z t\bar t$ vertex, and has  
right-handed contribution to $tbW$. At the 1-loop level, 
it contributes to $\epsilon_1$ and to $Zb \bar b$ in a 
left-handed way, \cite{bb}.  Present constraints from
$Z$-peak are rather marginal. In the linear colliders, the 
dominant effect should come from $A_{FB, pol}(e^-e^+ \to t \bar
t)$. \par

$\O_{tW\Phi}$:\\
It produces genuine tree level magnetic type $\sigma_{\mu \nu}$
couplings to the $\gamma t\bar t$, $ Z t\bar t$ and $tbW$
vertices. The $tbW$ vertex has the additional characteristic
that it only involves the left-handed $b_L$-field. $\O_{tW\Phi}$ 
is presently constrained only by its 1-loop contribution to 
$\epsilon_3$ \cite{bb}.
The observability limit in the linear colliders arises 
from the integrated production cross section, the top
decay width and the  decay asymmetries; 
(mainly the $D^2_{FB}$). This observability limit will supply
only  a very minor improvement to the present one from $\epsilon_3$.
This
operator could further be checked by looking for an enhancement in 
the decay $t\to WZb$, with the  $Z$ decaying into lepton 
pairs \cite{Heinson}. It can also contribute to $t \to WHb$,
provided $H$ is sufficiently light. \par

$\O_{tB\Phi}$:\\
It produces similar tree-level effects to the $\gamma t\bar t$ 
and $ Z t\bar t$ vertices, but no contribution to  $tbW$.
The effect in the integrated cross section is similar to the
one due to $\O_{tW\Phi}$, but the effect on $\epsilon_3$
is weaker, leaving more chance for observability. \par
 
$\O_{tG\Phi}$:\\
At the 1-loop level, it produces genuine magnetic type 
$\sigma_{\mu \nu}$
couplings to the $\gamma t\bar t$, $ Z t\bar t$ and $tbW$
vertices. The $tbW$ vertex has the additional characteristic
that it only involves the left-handed $b_L$-field.
These properties are like those for $\O_{tW\Phi}$; but 
appearing  at 1-loop,  rather than at tree level.
As a result,  there is now no 
contribution to $\epsilon_3$ at 1-loop. 
Future constraints to $\O_{tG\Phi}$ from linear colliders
should arise from studies of
the integrated cross section, the decay width and
the top decay asymmetries.\par
 
\vspace{0.5cm}
 
\underline {Bosonic operators.}\\
The effects of these operators on the $\gamma t \bar t$, 
$Z t \bar t$ or $tbW$ vertices,  arise only  at the 1-loop level.\par

$\O_W$:\\
It contributes only to the left-handed $\gamma  t\bar t$ 
and $Z t \bar t$ form factors.
A visible effect from them at a linear collider
could appear in $H_{FB}$ and in the cross section. 
The needed value of the coupling falls just below
the visibility domain of LEP2.  No effect is
generated in the decay.\par

$\O_{W\Phi}$: \\
It contributes to the vector and axial form-factors for the
$\gamma t \bar t$ and $Zt \bar t$ vertices. For the $tbW$
vertex,  $\O_{W\Phi}$
creates a left-handed and a  derivative coupling, such
that a left-handed $b_L$ field is only involved.
The visible effects from these couplings at a linear collider 
are similar to those expected from $\O_W$, and the same situation 
with respect to LEP2 is valid. The expected effects on $t\to bW$ 
seem to be below the observability level.\par
 
$\O_{B\Phi}$: \\
It induces the same type of couplings as $\O_{W\Phi}$.
The sensitivity in $H_{FB}$ is now somewhat better though;
so that there exists the possibility of a visible effect from the
$\gamma t \bar t$ and $Z t\bar t$ vertices, which  will not 
be already excluded by LEP2. The effects
on the $t\to bW$ decay are still unobservable.\par
 
$\O_{UW}$: \\
As in the $\O_{tG\Phi}$ case, it produces genuine magnetic type 
$\sigma_{\mu \nu}$ type
couplings to the $\gamma t\bar t$, $ Z t\bar t$ and $tbW$
vertices. The $tbW$ vertex has the additional characteristic
that it only involves a left-handed $b_L$-field. Note, that these
same properties also arise in the $\O_{tW\Phi}$ case, where
they are induced at the tree-level though. Another thing to note
is that $\O_{UW}$ is very mildly constrained by $Z$-peak physics, which
is also valid for $\O_{tG\Phi}$, but not true for $\O_{tW\Phi}$.
A most distinctive signature discriminating $\O_{UW}$ from the
other two operators, may be obtained by
studying $e^-e^+ \to ZH,~\gamma H$ \cite{HZ}. \par

$\O_{UB}$: \\
As far as the $\gamma t \bar t$ and $Z t \bar t$ vertices 
are concerned, the results are similar to the $\O_{UW}$ ones,
but their ratio is  different. 
For $\O_{UB}$ we have  $d^Z_j/d^\gamma_j=-2\swd$, while in  the 
$\O_{UW}$ case we have instead $d^Z_j/d^\gamma_j=-2 \cwd$.
No effect appears in the $t\to bW$ decay.\par
 
$\O_{\Phi 2}$: \\
This operator produces a purely axial  $Z t\bar t$ vertex, 
and a left-handed $tbW$ one. There is no $\gamma t \bar t$
vertex induced. Visible effects from $Zt \bar t$
could be obtained by looking at $H_{FB}$ 
and the integrated cross section. The study of 
$\Gamma (t\to bW) $ should also help. 
Like the two previous operators, it could however be more 
strongly constrained by direct Higgs
production.

\section{Panorama of residual NP effects in the heavy
quark and bosonic sectors}
\hspace{0.5cm}
We have considered the possibility of anomalous top quark couplings
induced by residual NP effects, described by twenty $dim=6$ gauge
invariant operators. ~Fourteen of them involve the top quark, and
the other six are purely bosonic. The couplings of these
operators are
associated to a NP scale through unitarity relations.\par 

We have computed the effects of these operators in
$e^+e^-\to t\bar t$ and $t\to bW$ decay. 
For $e^+e^-\to t\bar t$, these effects are described in terms of
six independent form factors
for the general $\gamma t\bar t$ and $Z t\bar t$ vertices.
Correspondingly,  the $t\to bW$ decay is described in terms 
of four couplings denoted as $d^W_j$. The top quark 
density matrix can thus be expressed
in terms of these form
factors and couplings. We have shown how one can 
analyze this density 
matrix in order to get information on the possible forms 
of NP induced by the various operators. 
The extra information brought by polarized $e^{\pm}$
beams, is also considered.\par

Thus, in addition to the integrated unpolarized cross
section and the L-R asymmetry, it is possible for polarized
beams to construct six different
forward-backward asymmetries which should allow to disentangle the
effects of the six form factors $d^\gamma_j$, $d^Z_j$. 
It is more difficult to disentangle
the NP effects on the $t\to bW$ decay, as no accurate
measurement of $\Gamma (t\to bW)$ is expected, 
and only one particular combination of decay
couplings can easily be measured through an asymmetry with
respect to the final lepton, in a semileptonic top 
decay\footnote{The same is also true if a hadronic mode is
considered.}. \par

The consequences of the NP operators on the above form factors
and $d^W_j$ couplings were calculated to
first order in the NP couplings. The calculation was done to the
tree level, whenever this gave a non-vanishing contribution.
In case there was no such contribution, we performed a
calculation at the 1-loop level, 
keeping only  the leading-log 
$\mt$-enhanced part. Numerical illustrations have been given for 
the various observables, which 
reflect the specific properties of each operator. We have
then established the corresponding observability limits for each
operator in terms of the associated coupling constant and
identified the related NP scale. 
The results can be summarized as follows.\par

\underline{ 4-quark operators}:\\
Among the seven 4-quark operators, four of them  $\O_{qt}$,
$\O^{(8)}_{qt}$, $\O_{tt}$, $\O_{tb}$, could give sizeable
1-loop effects in $e^+e^-\to t\bar t$. $\O_{tt}$ is not 
constrained by $Z$ peak physics, which means that $e^+e^-\to t\bar t$
would provide a completely new test. The three other operators  
produce 1-loop effects  also in $Z\to b\bar b$, which have been 
studied in \cite{bb}. The departure from SM
presently observed in $\Gamma(Z\to b\bar b)$, if attributed to one of
these three operators, could produce effects which should be
easily visible in the $\O_{tb}$ case,
but would only be marginally observable for  $\O_{qt}$ and
$\O^{(8)}_{qt}$.\par

Concerning the remaining three 4-quark operators, we note
that $\O_{qq}$ and $\O_{qq}^{(8)}$ are constrained essentially
only by $t \to b W$ \cite{bb}, while $\O_{tb}^{(8)}$ is not sensitive
to $Z$-peak physics or $e^-e^+ \to t \bar t$, $t \to b
W$.  
 
\underline{2-quark operators}:\\
Four of the seven 2-quark operators, $\O_{t2}$, $\O_{D t}$, 
$\O_{tW\Phi}$ and $\O_{tB\Phi}$, produce  tree-level effects
in $e^+e^-\to t\bar t$, while $\O_{t3}$ produces a tree-level
effect only in  $t \to bW$. $\O_{tG\Phi}$ contributes at 1-loop level
to both
production and decay. However the above  four operators
 $\O_{t2}$, $\O_{D t}$, $\O_{tW\Phi}$, $\O_{tB\Phi}$
generate also some $\epsilon_j$ contribution to 
Z-peak physics. This seems to already
exclude an observable effect from 
$\O_{t2}$ and $\O_{D t}$, but leaves some range for
observability to $\O_{tW\Phi}$ and $\O_{tB\Phi}$.\par
 
Concerning the rest of the 2-quark operators, we remark that
$\O_{t3}$ and $\O_{tG\Phi}$ are
presently unconstrained; so $e^-e^+ \to t \bar t$ and $t \to b W
\to b l^+ \nu$ would provided genuine new tests of them.
$\O_{t1}$ is not observable through these processes though, and
its study requires $t\bar t H$ production \cite{tth}.\par

\underline{Bosonic operators}:\\
All six bosonic operators contribute at 1-loop to $e^+e^-\to
t\bar t$. Moreover, 
$\O_{W\Phi}$, $\O_{UW}$ and $\O_{\Phi 2}$ could also significantly
contribute to $t\to bW$. The operators $\O_{W}$, 
$\O_{W\Phi}$ and $\O_{B\Phi}$ should have an observability level which
would not be excluded by LEP2. Notice that $\O_{W\Phi}$ has a 1-loop 
effect in $Z\to b\bar b$ which could explain the observed
anomaly there \cite{bb1}. In this case large effects should be 
observed in
$t\bar t$ production. However a direct study of these operators in  
$e^+e^-\to W^+W^-$ at NLC should be even more stringent.\par

The three other bosonic operators involving Higgs field are 
almost unconstrained at present. So nothing excludes
their appearance. However if the Higgs mass is low enough to allow for
$e^+e^-\to HZ$ or $e^+e^-\to H\gamma$ at LEP2 and/or at
NLC, then these processes would improve the sensitivity limits
on these operators by 2 orders of magnitude.\par

\vspace{0.5cm}

In conclusion the process $e^+e^-\to t\bar t$ should bring essential
~information on residual NP effects affecting the heavy quark
sector as well as the bosonic (gauge and scalar) sector. Its main
interest is that it provides
direct tests of the presence of genuine operators involving the third
family of quarks. It could give hints about the origin of the anomalies
recently observed in the $Zb\bar b$ couplings.

\vspace{1cm}
 
\underline{Acknowledgements}: We thank Claudio Verzegnassi for
having drawn our attention to the $m^2_t$-enhanced SM contributions
and to the bosonic NP contributions to the $t\bar t$ production
process.\par

\newpage

\renewcommand{\theequation}{A.\arabic{equation}}
\setcounter{equation}{0}
\setcounter{section}{0}

{\large \bf Appendix A : New physics vertices generated by the
effective lagrangian}\par
\vspace{0.5cm} 
It is easy to see that the leading-log $\mtd$-~enhanced
NP contributions to $e^-e^+\to t \bar t$ (up to the 1-loop
order), come exclusively from
vertex  diagrams for the $\gamma \to t \bar t$ and $ Z \to 
t\bar t$ vertices and from self-energies. The same is of course true
for the diagrams affecting $t \to bW$. Thus, box diagrams never
appear. Below we enumerate these
contributions for the various operators.\par

\noindent
\underline {Four-quark operators}:\\
They contribute to the vertices $\gamma t\bar t$, $Z t\bar t$, 
 $tbW$ through 1-loop diagrams involving a 4-quark interaction. 
This interaction can be
read off the following extended expressions. Thus, 
\bq
\O_{qt}=(\bar t_L t_R)(\bar t_R t_L)+(\bar b_L t_R)(\bar t_R b_L)
\eq
contributes through  $t$ and  $b$ loops to the 
$\gamma t\bar t$ and $Z t\bar t$ vertices, but not to the 
$t bW$ one.
\bq
\O^{(8)}_{qt}=(\bar t_L \overrightarrow\lambda t_R).
(\bar t_R \overrightarrow\lambda t_L)
+(\bar b_L \overrightarrow\lambda t_R).
(\bar t_R \overrightarrow\lambda b_L)
\eq
contributes through  a $b$ loop only to the 
$\gamma t\bar t$, $Z t\bar t$ vertices, but not to the 
$t bW$ one.
\bq
\O_{tt}={1\over2}\, (\bar t_R\gamma_{\mu} t_R)
(\bar t_R\gamma^{\mu} t_R)
\eq
contributes through
$t$-loop to $\gamma t\bar t$, $ Z  t\bar t$
but not to $t bW$.
\bq
\O_{tb}= (\bar t_R \gamma_{\mu} t_R)
(\bar b_R\gamma^{\mu} b_R)
\eq
contributes through a
$b$-loop to $\gamma t\bar t$, $ Z  t\bar t$
but not to $t\to bW$.
\bq
\O^{(8)}_{tb}= (\bar t_R\gamma_{\mu}\overrightarrow\lambda t_R)
(\bar b_R\gamma^{\mu} \overrightarrow\lambda b_R)
\eq
gives no contribution.
\bq
\O_{qq}=(\bar t_R t_L)(\bar b_R b_L)+(\bar t_L t_R)(\bar b_L b_R)
-(\bar t_R b_L)(\bar b_R t_L)-(\bar b_L t_R)(\bar t_L b_R)
\eq
contributes to the $tbW$ vertex, but not to $\gamma t\bar t$, $
Z t\bar t$. Finally, the
\bqa
\O^{(8)}_{qq} &= &(\bar t_R \overrightarrow\lambda t_L).
(\bar b_R \overrightarrow\lambda b_L)
+(\bar t_L \overrightarrow\lambda t_R).
(\bar b_L \overrightarrow\lambda b_R) \nonumber \\
&&-(\bar t_R \overrightarrow\lambda b_L).
(\bar b_R \overrightarrow\lambda t_L)
-(\bar b_L \overrightarrow\lambda t_R).
(\bar t_L \overrightarrow\lambda b_R)
\eqa
contributions are obtained from the $\O_{qq}$ ones, by
multiplying by the factor ${16\over3}$.\par

\noindent
\underline {Two-quark operators}:\par
Some of the operators in this class contribute already 
at the tree level, while others only at the 1-loop level. 
The later contributions arise from 
triangle  diagrams for the  $\gamma t \bar t$,
$Z t\bar t$ and $t bW$ vertices, as well as from 
fermion self-energy ones. 
When an operator contributes at
tree level, we do not care about its 1-loop contributions.\par

For the operator $\O_{t1}$, (after subtracting irrelevant
contributions to the top mass), we get 
\bq
\O_{t1}=[\chi^+\chi^- + vH + {1\over2}(\chi^3\chi^3  +H^2)]
[{v+H\over\sqrt2}(\bar t t)+{i\over\sqrt2}\chi^3(\bar t\gamma^5 t)
+i\chi^-(\bar b_L t_R) -i\chi^+(\bar t_R b_L)], 
\eq
which gives no contribution to the amplitudes  we
are interested. For the $\gamma , Z \to t \bar t$ amplitudes, this
comes about from the ~cancellation of the contributions from the 
vertex triangles involving ($ttH$) and ($tH\chi^3$) exchanges,  
and the ($tH$)-self-energy; while for the $t \to bW$ decay, 
the sum of the ($tH\chi^+$) triangle and the ($tH$)-self-energy 
vanishes.
The operator
\bqa
\O_{t2}&=&i(\bar t_L\gamma^{\mu}t_R)
\Big\{(\chi^-\partial_\mu\chi^+ - \partial_\mu\chi^-\chi^+)
+g(v+H)(\chi^-W^+_{\mu}-\chi^+W^-_{\mu}) \nonumber\\
&&-ig\chi^3(\chi^-W^+_{\mu} +\chi^+W^-_{\mu}) 
+ig_Z(1-2s^2_W)Z_{\mu}\chi^+\chi^-
+2ieA_{\mu}\chi^+\chi^- \nonumber\\
&&-i{g_Z\over2}Z_{\mu}[(v+H)^2+\chi^3\chi^3] +i\chi^3\partial_\mu H
-i(v+H)\partial_\mu\chi^3 \Big \}
\eqa
contributes at tree level to $t \bar t$ production.
\bqa
\O_{t3}&=&i(\bar t_R\gamma^{\mu}b_R) 
\Big\{ {i\over\sqrt2}[(v+H-i\chi^3)
\partial_\mu\chi^+ -\chi^+\partial_\mu(H-i\chi^3)] 
+{ig\over\sqrt2}\chi^+W^-_{\mu} \chi^+ \nonumber\\
&&+{ig\over2\sqrt2}W^+_{\mu}(v+H-i\chi^3)^2
-{g_Zc^2_W\over\sqrt2}(v+H-i\chi^3)Z_{\mu}\chi^+
\nonumber \\
&& -~{e\over\sqrt2} (v+H-i\chi^3)A_{\mu}\chi^+\Big \} \nonumber\\
&&-i(\bar b_R\gamma^{\mu}t_R)\Big \{{-i\over\sqrt2}[(v+H+i\chi^3)
\partial_\mu\chi^- -\chi^-\partial_\mu(H+i\chi^3)] 
-{ig\over\sqrt2}\chi^-W^+_{\mu} \chi^- \nonumber\\
&&-{ig\over2\sqrt2}W^-_{\mu}(v+H+i\chi^3)^2
-{g_Zc^2_W\over\sqrt2}(v+H+i\chi^3)Z_{\mu}\chi^-
\nonumber \\
&&-~{e\over\sqrt2} (v+H+i\chi^3)A_{\mu}\chi^- \Big \}
\eqa
has no effect on $t\bar t$  production, but contributes at tree 
level to the $t \to bW$ decay.
\bqa
&&\O_{D t}=\bar t_L\left [ \partial_\mu+ig'{2\over3}(-s_W Z_{\mu}+c_W
A_{\mu}) +i\frac{g_s}{2}\overrightarrow \lambda \cdot 
\overrightarrow G_\mu \right ] t_R 
 \cdot \Big [{1\over\sqrt2}\partial_\mu(H+i\chi^3)
\nonumber \\
&&+{ig\over2\sqrt{2}\cw}
(v+H+i\chi^3)Z_{\mu}-{g\over\sqrt2}W^+_{\mu}\chi^-\Big ] \nonumber\\
&&+ \bar b_L\left [\partial_\mu+ig'{2\over3}(-s_W Z_{\mu}+c_W
A_{\mu}) +i\frac{g_s}{2}\overrightarrow \lambda \cdot 
\overrightarrow G_\mu \right  ] t_R
\cdot \Big [i\partial_\mu\chi^- \nonumber \\
&&+{g(1-2s^2_W)\over2\cw}
Z_{\mu}\chi^-+eA_{\mu}\chi^-+{ig\over\sqrt2}W^-_{\mu}{v+H+i\chi^3\over
\sqrt2}\Big ] \nonumber\\
&&+\bar t_R \left [ \overleftarrow \partial_\mu
-ig'{2\over3}(-s_W Z_{\mu}+c_W
A_{\mu})-i\frac{g_s}{2}\overrightarrow \lambda \cdot 
\overrightarrow G_\mu \right  ] t_L \cdot
\Big [{1\over\sqrt2}\partial_\mu(H-i\chi^3)
\nonumber \\
&&-{ig\over2\sqrt{2}\cw}
(v+H-i\chi^3)Z_{\mu}-{g\over\sqrt2}W^-_{\mu}\chi^+\Big ] +\nonumber\\
&&~ \bar t_R \left [\overleftarrow \partial_\mu
-ig'{2\over3}(-s_W Z_{\mu}+c_W
A_{\mu}) -i\frac{g_s}{2}\overrightarrow \lambda \cdot 
\overrightarrow G_\mu \right ] b_L \cdot
\Big [-i\partial_\mu\chi^+  \nonumber \\ 
&&+{g(1-2s^2_W)\over2\cw}
Z_{\mu}\chi^+ +eA_{\mu}\chi^+
-{ig\over\sqrt2}W^+_{\mu}{v+H-i\chi^3)\over\sqrt2} \Big ]
\eqa
\noindent
contributes at tree level to both production and decay.
The same is true for
\bqa
\O_{tW\Phi}&=& (c_W Z_{\mu\nu}+s_W A_{\mu\nu}) 
\Big\{{1\over\sqrt2}(\bar t 
\sigma^{\mu\nu} t)(v+H) 
+{i\over\sqrt2}(\bar t \sigma^{\mu\nu}\gamma^5
t)\chi^3-i(\bar b_L \sigma^{\mu\nu} t_R)\chi^- \nonumber\\
&& +i(\bar t_R \sigma^{\mu\nu}
b_L) \Big \}\chi^+ 
+i\sqrt2(\bar t_L \sigma^{\mu\nu} t_R)W^+_{\mu\nu}\chi^-
 -i\sqrt2(\bar t_R \sigma^{\mu\nu} t_L)W^-_{\mu\nu}\chi^+ \nonumber\\
&&+(\bar b_L \sigma^{\mu\nu} t_R)W^-_{\mu\nu}(v+H+i\chi^3)
+(\bar t_R \sigma^{\mu\nu} b_L)W^+_{\mu\nu}(v+H-i\chi^3)
\ ,
\eqa
\noindent
while
\bqa
\O_{tB\Phi}&=&(-s_W Z_{\mu\nu}+c_W A_{\mu\nu})\Big 
\{{1\over\sqrt2}(\bar t 
\sigma^{\mu\nu} t)(v+H)\nonumber\\
&&+{i\over\sqrt2}(\bar t \sigma^{\mu\nu}\gamma^5
t)\chi^3+i(\bar b_L \sigma^{\mu\nu} t_R)\chi^- 
-i(\bar t_R \sigma^{\mu\nu}
b_L)\chi^+\Big \}
\eqa
\noindent
contributes at tree level only to $t\bar t$ production.
Finally
\bqa
\O_{tG\Phi}&=&G^a_{\mu\nu}\Big \{{1\over\sqrt2}(\bar t 
\sigma^{\mu\nu}\lambda^a t)(v+H)+{i\over\sqrt2}(\bar t \sigma^{\mu\nu}
\lambda^a\gamma^5
t)\chi^3\nonumber\\
&&+i(\bar b_L \sigma^{\mu\nu}\lambda^a t_R)\chi^- 
-i(\bar t_R \sigma^{\mu\nu}\lambda^a
b_L)\chi^+\Big \}
\eqa
\noindent
contributes at 1-loop to production through the ($ttg$) triangle and the 
($tg$) self-energy; and to $t\to bW$ decay through the 
($tbg$) triangle and the ($tg$) self-energy.\par
 
\noindent 
\underline {Bosonic operators}:\\
Contributions in this class arise only at the 1-loop level, through
triangle  and self-energies diagrams. Thus,
$\O_W$ (see (17)) 
contributes to $e^-e^+ \to t\bar t$  production through the ($WWb$)
triangle\footnote{$\O_W$ does not produce
$m_t^2$ terms but it must be taken into consideration
since the contribution is proportional to s which is
larger than $4m_t^2$ for the process under consideration.},
but gives no contribution to $t\to bW$, since 
the sum of the $m^2_t$-enhanced parts of the 
($tW\gamma$) and ($tWZ$) triangles vanishes. The operator
$\O_{W\Phi}$ (see  (18))
contributes to production through the ($tH\chi^3$) and
($\chi^+\chi^- b$)triangles; and to 
decay through the ($t\chi^+\gamma$), ($t\chi^+ Z$), ($b\chi^+\gamma$)
and ($b\chi^+ Z$) triangles. In a similar way
$\O_{B\Phi}$ (see (19)) 
contributes to production through the ($tH\chi^3$) and
($\chi^+\chi^- b$) triangles,  and to 
decay through the ($t\chi^+\gamma$), ($t\chi^+ Z$), ($b\chi^+\gamma$)
and ($b\chi^+ Z$) triangles. The operator
$\O_{UW}$ (see (20)) 
contributes to production through the ($tH\gamma$) and ($tHZ$)
triangles and to 
decay through ($tHW$). Correspondingly, $\O_{UB}$ (see (21))
contributes to production through the ($tH\gamma$) and ($tHZ$)
triangles, (like in the $\O_{UW}$ case), but 
gives no contribution to $t\to bW$ decay. Finally 
$\O_{\Phi 2}$ (see (22)) induces a renormalization of the
physical Higgs field at the tree level. This, in turn, gives
contributions to production, through the
$ttH$-triangle and the $tH$-self-energy, and to decay through
the $tH$-self-energy.

\newpage

\renewcommand{\theequation}{B.\arabic{equation}}
\setcounter{equation}{0}
\setcounter{section}{0}

{\large \bf Appendix B : Top decay distributions}\par
\vspace{0.5cm}
As discussed in Section 4, it is convenient to express 
the 3-body phase space 
$d\Phi_3(bl\nu_l)$ in terms of the
Euler angles determining the $t$-decay plane. 
We start from the process $e^-(k) e^+(k') \to
t (p) \bar t(p') $ in the center of mass frame,
where the momenta are indicated in parentheses; and by 
$\theta$ we denote the $(e^-,t)$ scattering angle. 
The $t$-frame is defined with its z-axis
along  the top-quark momentum. The 
x-axis is taken in the $(t~\bar t)$ production plane, so that
the y-axis is perpendicular to 
it and along the
direction of   $\kvec \times \pvec$. 
In order to describe the decay-plane of the process 
 $t \to b(p_b) l^+(p_l) \nu_l(p_\nu)$, in the 
$t$-frame  (with the momenta indicated in parentheses), 
we define the Euler rotation
\bqa
&& R_{\varphi_1 \vartheta_1 \psi_1}  = \nonumber \\
&& \left (\matrix{\cos\varphi_1 & -\sin\varphi_1 & 0 \cr
               \sin\varphi_1 & \cos\varphi_1 & 0 \cr
               0 & 0 &1 } \right ) 
\left (\matrix{\cos\vartheta_1 & 0 & \sin\vartheta_1  \cr
                     0 & 1 & 0                      \cr
      -\sin\vartheta_1 & 0 & \cos\vartheta_1 } \right ) 
        \left (\matrix{\cos\psi_1 & -\sin\psi_1 & 0 \cr
               \sin\psi_1 & \cos\psi_1 & 0 \cr
               0 & 0 &1 } \right ) \ \ ,
\eqa
where  $(\varphi_1, \vartheta_1, \psi_1)$
satisfy $ 0\leq \varphi_1, \psi_1 < 2\pi$ , $ 0 \leq
\vartheta_1 \leq \pi $. The meaning of these angles
is given by remarking that the 
normal to the $t$-decay plane, with its ~orientation defined by  
($\pvec_b \times \pvec_l $), is given by
\bq
\hat{n} = R_{\varphi_1 \vartheta_1 \psi_1}\cdot
\left (\matrix {0 \cr 0 \cr 1} \right ) =
\left (\matrix {\sin\vartheta_1 \cos\varphi_1 \cr 
       \sin\vartheta_1 \sin\varphi_1 \cr 
       \cos\vartheta_1} \right ) \ .
\eq
Thus, $\vartheta_1, \varphi_1$ determine the $\hat{n}$ 
orientation, 
while the b quark momentum in the $t$-rest frame is determined
from $\psi_1$  through the relation
\bq
\overrightarrow p_b = |\overrightarrow p_b|
R_{\varphi_1 \vartheta_1 \psi_1}\cdot
\left (\matrix {1 \cr 0 \cr 0} \right ) =|\overrightarrow p_b|
\left (\matrix { 
 \cos\varphi_1 \cos\vartheta_1 \cos\psi_1 - \sin\varphi_1 \sin\psi_1 \cr 
 \sin\varphi_1\cos\vartheta_1\cos\psi_1 +\cos\varphi_1\sin \psi_1       \cr 
      - \sin\vartheta_1 \cos\psi_1} \right ) \ .
\eq
The corresponding expression for the $l^+$ momentum is obtained
from (B.3) by substituting $\pvec_b \to \pvec_l$ and $\psi_1 \to
\psi_1 +y_{12}$, where $y_{12}$ is the angle between the $b$ and
$l^+$ momenta (in the $t$ rest frame). To summarize, it is
worthwhile to remark that the above Euler rotation moves the
z-axis of the $t$-frame along the normal to the $t$-decay
plane; while the x-axis is brought along the $\pvec_b$ momentum,
which of course lies within the decay plane. \par

Finally $\theta_l$ is the
angle between the lepton momentum and the top momentum in the W-rest
frame, and it is related to the $l^+$ energy in the $t$-frame by
\bq
E_l=|\overrightarrow p_l| =~\frac{\mtd +\mwd -\cos\theta_l
(\mtd-\mwd)}{4 \mt} \ \ ,
\eq
where the $(b,l^+)$ masses are neglected.
Using these Euler angles, we obtain 
\bq
\delta((p_l+p_\nu)^2-\mwd) 
d\Phi_3(bl\nu_l) \Rightarrow
{(m^2_t-M^2_W)\over64m^2_t(2\pi)^9}d\varphi_1d\cos
\vartheta_1d\psi_1 d\cos\theta_l \ \ ,
\eq
for the 3-body phase space in the case where the 
$l\nu$-pair is at the $W$-mass shell \cite{PDG}.\par

The general expression 
of the differential cross section for
$e^+e^-\to t\bar t$ with $ t\to bW \to b l \nu_l $ and linearly
polarized $[L (R)~ ~e^-]$ and $[R(L)~ ~e^+]$ beams, is 
written (compare (55,56)) as
\bq
\frac{d\sigma^{L, R}}{d\cos\theta d\varphi_1 d\cos\vartheta_1
d\psi_1 d\cos\theta_l}
=\left ({3\beta_t\over 32 (2\pi)^5 s}\right )
\frac{ G^2_F M^3_W}{\Gamma_W\Gamma_t\mt} \left 
(\frac {m^2_t-M^2_W}{4\mt}\right )^2
\rho^{L, R}_{\tau_1 \tau_2}\cdot \R_{\tau_1 \tau_2} \ ,
\eq
where $\beta_t=(1-{4m^2_t\over s})^{1/2}$. In (B.6),
$\rho^{L,R}$ is
the top density matrix  defined in (50),
while $\R$ is related to the top decay matrix 
$t_{\tau_1 \tau_2}$ introduced in (56,58) by  
\bq
\frac{t_{\tau_1 \tau_2}}{d\varphi_1 d\cos\vartheta_1
d\psi_1 d\cos\theta_l}=~\frac{G^2_F\mw^3}{2
(2\pi)^4\Gamma_W\Gamma_t \mt}\left 
(\frac{\mtd -\mwd}{4\mt} \right )^2 \R_{\tau_1 \tau_2} \ .
\eq\par 
 
The $\rho$ matrix depends only on the 
$e^-e^+ \to  t\bar t$-production and the angle 
$\theta$, (see (50)); while $\R$ depends on the three Euler 
angles $\varphi_1$,
$\vartheta_1$, $\psi_1$, (defined in (B.1)) and 
on the $d^W_j$ couplings of (\ref{eq:dw}) and the angle 
$\theta_l$.
To simplify the expression for $\R$, we only keep
 terms  linear in the NP and the 1-loop SM contributions to
the couplings $d^W_j$, defining $\bar d^W_j\equiv 
d^{W,SM1}_j + d^{W,NP}_j$ . We thus get   
\bqa
\rho^{L, R}_{\tau_1 \tau_2}
\cdot \R_{\tau_1 \tau_2}&=
&\frac{1}{2}~
 (\rho_{++}+\rho_{--})^{L, R}(\R_{++}+\R_{--})+
~ \frac{1}{2} ~
(\rho_{++}-\rho_{--})^{L, R}(\R_{++}-\R_{--})\nonumber\\
&&+~ \rho^{L, R}_{+-}(\R_{+-}+\R_{-+})\ \ ,
\eqa
where 
\bqa
\R_{++}+\R_{--}&=&(1+\bar d^W_1-\bar d^W_2)[M^2_W V_1+m^2_t V_2]
\nonumber\\
&&+\left ({m^2_t-M^2_W\over m_t}\right )(\bar d^W_3+ \bar d^W_4)
m^2_t V_2 \ ,
\eqa
\bqa
\R_{++}-\R_{--}&=&(1+\bar d^W_1-\bar d^W_2)(m^2_t V_4- 
M^2_W V_3+2m_t M_W V_5]
\nonumber\\
&&+\left ({m^2_t-M^2_W\over m_t}\right)
(\bar d^W_3+\bar d^W_4) [m^2_t V_4+m_t M_W V_5] \ ,
\eqa
\bqa
\R_{+-}+\R_{-+}&=&(1+\bar d^W_1-\bar d^W_2)
[M^2_W V_6-m^2_t V_7-2m_t M_W V_8]
\nonumber\\
&&-\left({m^2_t-M^2_W\over m_t}\right) 
(\bar d^W_3+\bar d^W_4)[m^2_t V_7+m_t M_W V_8] \ ,
\eqa
and  
\bq
V_1=(1+\cos\theta_l)^2 \ \ , \ \  V_2=\sin^2\theta_l \ ,
\eq
\bq
V_3=(1+\cos\theta_l)^2\sin\vartheta_1\cos\psi_1 \ \ , \ \
V_4=\sin^2\theta_l \sin\vartheta_1\cos\psi_1 \ ,
\eq
\bq
V_5=(1+\cos\theta_l)\sin\theta_l \sin\vartheta_1\sin\psi_1
\ ,
\eq
\bq
V_6=(1+\cos\theta_l)^2(\cos\varphi_1\cos\vartheta_1\cos\psi_1-
\sin\varphi_1\sin\psi_1) \ ,
\eq
\bq
V_7=\sin^2\theta_l(\cos\varphi_1 \cos\vartheta_1 \cos\psi_1
  -\sin\varphi_1\sin\psi_1)\ ,
\eq
\bq
V_8=\sin\theta_l(1+\cos\theta_l)(\cos\varphi_1 \cos\vartheta_1
  \sin\psi_1 +\sin\varphi_1 \cos\psi_1) \ .
\eq\par

Using the angular dependence in (B.9-17) and constructing appropriate
averages over $\varphi_1$, $\vartheta_1$ and  $\psi_1$, 
it is possible to project quantities proportional to the $\rho$
factors in each of the three terms of the r.h.s. in (B.8). 
More ~explicitly these ~quantities consist of products of the
corresponding $\rho$ elements, and of  
functions of $\theta_l$. Remember that the $\rho$ elements
depend only on $\theta$ and the NP couplings for $\gamma t \bar
t$, $Zt \bar t$. Thus, the subsequent construction
of forward-backward asymmetries with respect
to either $\theta$ or $\theta_l$ respectively, allows the isolation of
either the $\rho$ factor or of the corresponding combination of
the $\bar d^W_j$ couplings. 
To do this we first describe the $\rho$ elements   
entering the three terms in (B.8). For this, It is convenient to 
define for $i=1,2,3$ (compare (44))
\bq
d^L_i=d^{\gamma}_i+{1-2s^2_W\over4s^2_Wc^2_W}\chi d^Z_i \ \  , \ \ 
d^R_i=d^{\gamma}-{\chi\over2c^2_W}d^Z_i \ ,
\eq
where $\chi\equiv s/(s-M^2_Z)$ and the Z
width is neglected for $s=q^2 >4m^2_t$. We then have
\bqa
&&(\rho_{++}+\rho_{--})^{L,R}= e^4\sin^2\theta
\left({8 m^2_t\over s}\right )
\left [d^{L,R}_1-{2|\pvec|^2\over m_t} d^{L,R}_3 \right ]^2 \nonumber\\
&+&2e^4(1+\cos^2\theta)
\left [(d^{L,R}_1)^2+{4|\pvec|^2\over
s}(d^{L,R}_2)^2  \right ] 
\mp e^4\cos\theta \left ({16|\pvec |\over\sqrt{s}}
\right ) d^{L,R}_1d^{L,R}_2   ,
\eqa
\bqa
&&(\rho_{++}-\rho_{--})^{L,R}=e^4(1+\cos^2\theta)
\left ({8|\pvec| \over\sqrt{s}}\right )~
 d^{L,R}_1d^{L,R}_2 \nonumber\\
&\mp & 4e^4\cos\theta
\left [ (d^{L,R}_1)^2+{4|\pvec|^2\over s}(d^{L,R}_2)^2 
\right ] \ ,
\eqa
\bq
\rho^{L,R}_{+-}=e^4\sin\theta\left({4m_t\over\sqrt{s}}
\right)
\Big [d^{L,R}_1 - \frac{2|\pvec|^2}{\mt} d^{L,R}_3 \Big ]
\Big [d^{L,R}_1 - \frac{2|\pvec|}{\sqrt{s} }\cos\theta
d^{L,R}_2 \Big ]
\ \ .
\eq\par

For ~unpolarized $e^\mp$ beams, only the $(L+R)/2$ combination,
like \eg~ $d\sigma^{unpol}=(d\sigma^L+d\sigma^R)/2$ or 
$(\rho^L +\rho^R)/2$,
is measurable through forward-backward asymmetries. There are
three $\rho$ and $\R$ elements that can be studied this way.
If longitudinal electron beam polarization is available, we can 
also consider
the corresponding three ($\rho^L-\rho^R $) combinations and 
their forward-backward asymmetries.\par

Thus, by integrating both sides of (B.6) over 
$d\varphi_1 d\cos\vartheta_1 d\psi_1$,
the first term in the r.h.s. of (B.8) is projected. Integrating
also over
$\cos\theta_l$, and constructing the forward-backward asymmetry
with respect to the $t \bar t$ production angle $\theta$, allows
the study of the NP effects in $(\rho_{++} +\rho_{--})^{L,R}$.
This asymmetry is of course the usual forward-backward asymmetry
in the differential cross section for the top production
through $e^+e^-\to t\bar t$. We thus have 
\bqa
&&A_{FB}= \nonumber \\
&&{{3 \beta_t \over 2}[d^R_1d^R_2-d^L_1d^L_2]\over
(d^L_1)^2+(d^R_1)^2 + \beta_t^2 [ (d^L_2)^2+ (d^R_2)^2] 
+{2m^2_t\over s} [(d^L_1-{2|\pvec|^2\over m_t}d^L_3)^2+(d^R_1-
{2|\pvec|^2\over m_t}d^R_3)^2]} ,
\eqa
for the ~unpolarized case, while for the $L-R$ one we have
\bqa
&&A_{FB,pol}= \nonumber \\
&&{-~{3\beta_t \over 2}[d^R_1d^R_2+d^L_1d^L_2]\over
(d^L_1)^2-(d^R_1)^2 + \beta_t^2 [ (d^L_2)^2 - (d^R_2)^2] 
+{2m^2_t\over s} [(d^L_1-{2|\pvec|^2\over m_t}d^L_3)^2-(d^R_1-
{2|\pvec|^2\over m_t}d^R_3)^2]} .
\eqa\par

The preceding method for constructing the forward-backward 
asymmetry is just given in order to emphasize its 
similarity to the methods for constructing  the other
asymmetries bellow.
Consequently, by multiplying both sides of (B.6) by either
$\cos\psi_1$ or $\sin\psi_1$ and integrating over 
$d\varphi_1 d\cos\vartheta_1 d\psi_1$,
the second term in the r.h.s. of (B.8) is projected. Integrating
then over
$\cos\theta_l$, we construct the forward-backward asymmetry
with respect to $\theta$,
for the quantity $(\rho_{++} - \rho_{--})^{L,R}$
~controlling the angular distribution of the 
average helicity of the produced top-quark. Thus, in terms of the 
couplings defined in (\ref{eq:dgz}), the forward-backward 
asymmetry in the top-quark average helicity is
\bq
H_{FB}=-~{3 \Big \{ (d^L_1)^2-(d^R_1)^2 
+\beta_t^2 [ (d^L_2)^2 -(d^R_2)^2]\Big \}
\over 8 \beta_t [d^L_1d^L_2 +d^R_1d^R_2]} \ ,
\eq
for the  $e^\pm$ ~unpolarized case, while for the $L-R$ one we have 
\bq
H_{FB,pol}=-~{3 \Big \{ (d^L_1)^2+(d^R_1)^2 
+\beta_t^2 [ (d^L_2)^2 +(d^R_2)^2]\Big \}
\over 8 \beta_t [d^L_1d^L_2 -d^R_1d^R_2]} \ .
\eq

Finally the third term in the r.h,s, of (B.8) is projected by
multiplying both sides of (B.6) by quantities like any one of
\bq
\cos\psi_1\sin\varphi_1 \ \ , \ \
\sin\psi_1\cos\varphi_1\cos\vartheta_1 \ ,
\eq
\bq
\sin\psi_1\sin\varphi_1 \ \ , \ \
\cos\psi_1\cos\varphi_1\cos\vartheta_1\ , 
\eq 
and integrating  
over $d\varphi_1 d\cos\vartheta_1 d\psi_1$. The subsequent
integration over $\cos\theta_l$ allows  the construction of
the forward-backward asymmetry with respect to $\theta$ for
the quantity $\rho_{+-}^{L,R}$ ~controlling the angular
distribution of the  top average transverse polarization. 
Thus, for unpolarized $e^\mp$ beams, the forward-backward 
asymmetry in the top transverse polarization is obtained,
which is given by
\bq
T_{FB}=-{4 \beta_t \over3\pi }{\left [d^L_1d^L_2+d^R_1d^R_2
-\left ({2|\pvec|^2\over m_t}\right )
(d^L_3d^L_2 +d^R_3d^R_2) \right]\over
\left[(d^L_1)^2 -(d^R_1)^2
-\left( {2|\pvec|^2\over m_t} \right )(d^L_3 d^L_1 - d^R_3  d^R_1)
\right ]}\ ,
\eq
while for polarized beams the $L-R$ case gives
\bq
T_{FB,pol}={-4 \beta_t \over3\pi }{\left [d^L_1d^L_2-d^R_1d^R_2
-\left ({2|\pvec|^2\over m_t}\right )
(d^L_3d^L_2 -d^R_3d^R_2) \right]\over
\left[(d^L_1)^2 +(d^R_1)^2
-\left( {2|\pvec|^2\over m_t} \right )(d^L_3 d^L_1 + d^R_3  d^R_1)
\right ]}\ .
\eq\par

For any of the
~preceding three types of forward-backward asymmetries 
sensitive to the $t\bar t$
production couplings, we can construct corresponding 
asymmetries sensitive to
the decay couplings $\bar d^W_j$. This is done in all cases
by integrating at the last step over $\cos\theta$ 
(instead of over $\cos\theta_l$ done above) and
constructing the forward-backward asymmetry
with respect to $\theta_l$. As before, we always work to linear 
order in $\bar d^W_j$.\par

Thus, for the first case which led to (B.22,23) we get 
\bq
D^1_{FB}=\frac{3\mwd}{2(2\mwd+\mtd)}\Big [
1-\frac{\mt(\mtd-\mwd)}{2\mwd +\mtd}(\bar d^W_3+\bar d^W_4) \Big
] \ \ .
\eq\par

For the second case, we have already stated that the asymmetries 
(B.24,25) are obtained by using either   
$\cos\psi_1$ or $\sin\psi_1$ to project out the $\rho$ factor in
the second term in the r.h.s. of (B.8). 
For the $\theta_l$ asymmetry
though, these two projections give  different
asymmetries. Thus the asymmetry obtained through $\cos\psi_1$
is  
\bq
D^2_{FB}=\frac{- 3\mwd}{2(\mtd-2\mwd)}\Big [
1-\frac{\mt(\mtd-\mwd)}{\mtd-2\mwd}(\bar d^W_3+\bar d^W_4) \Big
] \ \ ,
\eq
while the one obtained from  $\sin\psi_1$ is independent of
$\bar d^W_J$ and equal to
\bq
D^3_{FB} = \frac{4}{3\pi} \ \ .
\eq
Finally in the third case, we get $D^2_{FB}$ for the asymmetry
obtained through the projector (B.27), and $D^3_{FB}$ for the
asymmetry obtained through (B.26).\par

To linear order in the NP couplings, all these ~asymmetries 
can be expressed as a product of a factor describing the SM
contribution, and another factor describing the NP correction. 
For this NP correction a tree level calculation is sufficient.
Any QCD  and 1-loop radiative corrections 
should in general be incorporated in the SM factor only.
The QCD corrections have to some extent been ~studied in
\cite{Kuhn, Lampe} and have been found rather small.
In any case this is something which we plan to do in the future.
It is also ~interesting to remark that while the production
asymmetries are sufficient to determine all $d^\gamma_j$
and $d^Z_j$ couplings even in the unpolarized case;
this is not possible  for the decay couplings. To
linear order in the NP top-decay couplings, the above 
asymmetries are only 
sensitive to the ~combination $\bar d^W_3+\bar d^W_4$. \par

Finally we should also remark that the case 
where the $t$-quark decays hadronically, while $\bar t \to
\bar b l^- \bar\nu $, is very similar. 
Thus, if the ~orientation of the
$\bar t$-rest frame is defined to be like the one obtained from
the $t$-frame by rotating it by $180^0$ around the perpendicular
to the $t\bar t$ production plane;  and if
the new Euler angles for the $\bar t$ decay plane are called 
$(\varphi_2,~ \vartheta_2,~ \psi_2)$, and $\theta_l$ is defined 
analogously; then all formulae in this Appendix remain the same,
except of (B.12-17) where we should replace
\bq
\varphi_1 ~,~ \psi_1 ~,~\vartheta_1 ~  \Rightarrow ~ \varphi_2 ~,~ \psi_2
~,~ -\vartheta_2  \  \ .
\eq 
This way, all forward-backward asymmetries remain formally
identical. Note though, that the definition of the top
production angle $\theta$ implies that (forward-backward) for 
$\bar t$ means that we should subtract as 
\bq
\mbox{Backward} (\bar t) -\mbox{Forward} (\bar
t) \ \ . 
\eq

\end{document}